\documentclass[journal,onecolumn]{IEEEtran}

\usepackage{datenumber}
\usepackage{amsmath} 
\usepackage{amssymb} 
\usepackage{amsfonts} 
\usepackage{verbatim}  
\usepackage{color} 
\usepackage{booktabs} 
\def\arraystretch{1.6}
\usepackage{nicefrac} 
\usepackage{multirow}
\usepackage{bigstrut}     
\usepackage{array}  
\usepackage{arydshln}  
\usepackage{esvect}  
\usepackage{pgfplots} \usetikzlibrary{plotmarks,patterns}
\usepackage{pgfplotstable}
 \usepackage{url}
 \usepackage{enumerate}
\usepackage{mathtools}
\usepackage{hhline} 
\usepackage{afterpage}

\newcommand{\Span}[1]{\mathsf{span} \!\left(  #1  \right)}
\newcommand{\RSpan}[1]{\mathsf{rspan} \left( #1 \right)}

\newcommand{\Rank}[1]{\mathsf{rank} \big( #1 \big)}
\newcommand{\Dim}[1]{\mathsf{dim} \left( #1 \right)}

\newcommand{\Bdiag}{\mathsf{bdiag}}
\newcommand{\bdiag}[1]{\Bdiag \left( #1 \right)}

\newcommand{\Lcm}[1]{\mathsf{lcm} \pare{#1}}
\newcommand{\Stack}{\mathsf{stack}}
\newcommand{\stack}[1]{\Stack \left( #1 \right)}

\newcommand{\pare}[1] {\left( #1 \right)} 

\newcolumntype{C}{>{$}c<{$}} 
\newcolumntype{L}{>{$}l<{$}}
\newcolumntype{R}{>{$}r<{$}}

\newcommand{\0}{\mathbf{0}}
\newcommand{\Ac}{\mathcal{G}}

\newcommand{\BS}[1]{\Tx{#1}}

\newcommand{\bbeta}{\overline{\beta}}

\newcommand{\Cmat}[2]{\!\in\mathbb{C}^{#1 \times #2}}

\newcommand{\dCJ}{d^{(\text{CJ})}}
\newcommand{\dHC}{d^{(\text{HC})}}
\newcommand{\dMAT}{d^{(\text{MAT})}}
\newcommand{\dTMAT}{d^{(\text{TMAT})}}
\newcommand{\duMAT}{d^{(\text{uMAT})}}

\newcommand{\din}{d^{(\text{in})}}

\newcommand{\enters}{\mathbb{Z}^+}

\newcommand{\Gdes}{\Gn_j^{\text{des}}}

\newcommand{\Gn}{\mathbf{\Omega}}

\newcommand{\Hn}{\mathbf{H}}
\newcommand{\hn}{\mathbf{h}}
\newcommand{\hns}[2]{\underline{\hn}_{#1}^{\left(#2\right)}} 
\newcommand{\Hns}[2]{\Hn_{#1}^{\left(#2\right)}}

\newcommand{\Rx}[1]{\text{UE}_{#1}}

\newcommand{\sigmas}[2]{{\boldsymbol{\sigma}}_{#1}^{(#2)}}
\newcommand{\Sigmas}[2]{\boldsymbol{\Sigma}_{#1}^{(#2)}}
\newcommand{\tng}[2]{\underline{\mathbf{m}}_{#1}^{(#2)}}
\newcommand{\Tc}{\mathcal{T}}

\newcommand{\Tx}[1]{\text{BS}_{#1}}
\newcommand{\Tn}{\mathbf{T}}
\newcommand{\Tns}[1]{\Tn_{#1}}
\newcommand{\UE}[1]{\Rx{#1}}
\newcommand{\un}{\mathbf{u}}

\newcommand{\Vn}{\mathbf{V}}

\newcommand{\Vns}[2]{\Vn_{#1}^{\left(#2\right)}}

\newcommand{\xn}{\mathbf{x}}
\newcommand{\Xn}{\mathbf{X}}
\newcommand{\Xc}{\mathcal{X}}
\newcommand{\yn}{\mathbf{y}}
\newcommand{\Yn}{\mathbf{Y}}

\newcommand{\Wn}{\mathbf{W}}

\newtheorem{theorem}{Theorem}
\newtheorem{lemma}{Lemma}

\newtheorem{corollary}{Corollary}

\newtheorem{conjecture}{Conjecture}
\IEEEoverridecommandlockouts
\renewcommand{\IEEEQED}{\IEEEQEDopen} 

\newcolumntype{C}{>{$}c<{$}} 
\newcolumntype{L}{>{$}l<{$}}
\newcolumntype{R}{>{$}r<{$}}

\definecolor{creme}{RGB}{255, 253, 208}

\pgfplotsset{
   plotoptsMAT/.style={color=cyan, thick, densely dotted, mark=|, mark size=1.5pt, mark options={solid}},
  plotoptsCoop2/.style={color=black, thick, densely dotted, mark=x, mark size=1.5pt, mark options={solid}},
   plotoptsCoop3/.style={color=black, thick, densely dotted, mark=triangle*,  mark options={solid}},
   plotoptsPropIn2/.style={color=blue, thick, mark=x, mark size=1.5pt, mark options={solid}},
   plotoptsPropIn3/.style={color=blue, thick, mark=triangle*},
   plotoptsPropIn4/.style={color=blue, thick, mark=square*},
   plotoptsPropIn5/.style={color=blue, thick, mark=pentagon*,  mark size=1.2pt, mark options={solid}},
   plotoptsCastanheiraC5/.style={color=green!60!black, thick, dashed, mark=pentagon*, 
   mark size=1.2pt,  mark options={solid}},
      plotoptsIC2/.style={color=red, thick, dashed, mark=x, mark size=1.5pt, mark options={solid}},
   plotoptsIC3/.style={color=red, thick, dashed, mark=triangle*,  mark options={solid}},
   plotoptsIC4/.style={color=red, thick, dashed, mark=square*,  mark options={solid}},
      plotoptsIC5/.style={color=red, thick, dashed, mark=pentagon*,  mark size=1.2pt,  mark options={solid}}
}
\tikzstyle{every pin}=[font=\footnotesize,pin distance=0.6cm,inner sep=1pt]

\begin{document}
\title{On the Degrees of Freedom of the MISO Interference Broadcast Channel with Delayed CSIT} 
\author{Marc~Torrellas,
        Adrian~Agustin,
        and~Josep~Vidal
\thanks{The authors are with the Signal Theory and Communications Department at the Universitat Polit\`ecnica de Catalunya (UPC), Barcelona \{marc.torrellas.socastro, adrian.agustin, josep.vidal\}@upc.edu. This work has been supported by projects 5G\&B RUNNER-UPC
TEC2016-77148-C2-1-R (AEI/FEDER, EU) and the Catalan Government
(2017 SGR 578-AGAUR)}}

 
\markboth{Transactions on Information Theory, \today}%
{Torrellas \MakeLowercase{\textit{et al.}}: Uncoupling the MAT scheme for the MISO Interference Broadcast Channel}

\maketitle

\begin{abstract}
The Maddah-Ali and Tse (MAT) scheme is a linear precoding strategy that exploits Interference Alignment and perfect, but delayed, channel state information at the transmitters (delayed CSIT), improving the degrees of freedom (DoF) that can be achieved for the broadcast channel (BC). Since its appearance, many works have extended the concept of Retrospective Interference Alignment (RIA) to other multi-user channel configurations. However, little is known about the broadcast channel with multiple cells, i.e. the interference broadcast channel (IBC). In this work, the DoF are studied for the 
$K$-user $C$-cell multiple-input single-output (MISO) IBC with delayed CSIT (with $K/C$ users per cell). 

We show that the straightforward application of the MAT scheme over the IBC fails because it requires all interference to be received from the same source. Hence, in this case not all the interference can be cancelled, thus blocking the decoding of the received messages. We call this phenomenon as \textit{interference coupling}, forcing to use the MAT scheme by serving just one cell at a time. In this work, we propose an extension, namely the uncoupled MAT scheme (uMAT), exploiting multiple cells, uncoupling the interference, and achieving the best known DoF inner bound for almost all settings. 
\end{abstract}


\section{Introduction}
%
%

\IEEEPARstart{I}{}nterference Alignment (IA) is a revolutionary technique that appeared just a decade ago  \cite{Maddah-Ali2008}\cite{CJ} for managing the signal dimensions (time, frequency, space) in pursuit of attaining the optimal multiplexing gain, a.k.a. the channel \textit{degrees of freedom} (DoF). These represent the scaling of channel capacity w.r.t. the signal-to-noise ratio (SNR) at the high SNR regime. While the capacity has been characterized for single-terminal scenarios, less is known for multi-terminal ones. Hence, characterizing the DoF sheds some light about how available channel state information (CSI), or the number of transmit or receive antennas, impact on system capacity. In this context, IA entailed a breakthrough since it allows providing \textit{each user half the cake} as compared to the single-user case, i.e. a total of\footnote{CJ denote here the initials of the original authors of the first Interference Alignment scheme for the Interference Channel: Viveck Cadambe and Syed Ali Jafar.}
\begin{IEEEeqnarray}{c}
\dCJ = \frac{Km}{2}
\label{eq:CJ}
\end{IEEEeqnarray}
are achieved over a network with $K$ transmitter-receivers pairs\footnote{This scenario could represent multiple cells with one user served per cell in the same frequency band.}, with each node equipped with $m$ antennas, a.k.a the $K$-user multiple-input multiple-output (MIMO) interference channel (IC). The concept consists on designing the transmitted signals in such a way that they are overlapped (or \textit{aligned}) at the non-intended receivers. Therefore, the interference lies on a reduced dimensional subspace, releasing some dimensions to allocate desired signals which can be retrieved by means of zero-forcing (ZF) receivers, see \cite{ZFBD}.

While ZF-based strategies suffice to characterize the DoF for the broadcast channel (BC), i.e. a downlink scenario with one base station (BS) serving to multiple user equipments (UEs), it revealed as a very useful tool for the study of other multi-user scenarios \cite{GenieTreeIBC}\cite{AlignmentChainsTrans}, 
such as the IC mentioned above. An extensive survey of IA applications can be found in \cite{IAtutorial2}. Of special interest for this work is \cite{GenieTreeIBC}, where IA is applied to the multi-cell BC, a.k.a. the interference BC (IBC), thus multiple BSs serving to multiple UEs each. In such scenario, the authors of \cite{ZFBD} proposed an extension of ZF, a.k.a. ZF block-diagonalization (ZF-BD) scheme, mimicking the results from the BC into the IBC. However, ZF-BD only applies when there are many more antennas at the BS than at the UE side. Hence, a combination of IA and ZF is required to exploit all signal dimensions. The conclusions of \cite{GenieTreeIBC} are twofold. On the one hand, for some settings the optimal DoF are obtained by interpreting the IBC as an IC where the fact that every BS knows all the messages intended to the UEs in its cell is not exploited. On the other hand, the rest of settings require applying an \textit{alignment graph}, an extension of the \textit{alignment chain} concept introduced in \cite{AlignmentChainsTrans} for the IC, and conveniently adapted to the IBC.

All IA-based and ZF-based schemes previously mentioned require perfect and current channel state information at the transmitters (CSIT), an assumption not always realistic in wireless cellular networks. For example, in frequency division duplexing systems, channel coefficients are usually estimated at the receivers by means of a training period, and then fed back to the transmitters, introducing delays and errors. The feedback error has been widely studied in the literature, and the main conclusion is that in order to preserve the DoF, the number of quantization bits should scale with the logarithm of the SNR \cite{Jindal}.  
On the other hand, it is usually assumed a block fading channel model, where channel remains constant in blocks of duration equal to the channel coherence time. When the feedback delay is longer than the coherence time, the available CSIT is completely outdated, and all strategies based on current CSIT are no longer effective. 
 
In this respect, Maddah-Ali and Tse (MAT) introduced in \cite{MAT} a new protocol framework where IA concepts can be exploited even when the CSIT is completely outdated, referred to as delayed CSIT. Indeed, they assume {\it perfect} delayed CSIT, which is now more feasible since during the time elapsed between transmissions, receivers can report channel coefficients with more resolution. The MAT scheme was the first application of IA concepts using delayed CSIT only. Originally proposed for the $K$-user multiple-input single-output (MISO) BC, the communication is carried out along $K$ phases for transmitting multiple symbols per user, achieving a total DoF sum given by\footnote{The second inequality uses an approximation for the summation of reciprocals of integers, see e.g. \cite{gradshteyn2007}.}
\begin{IEEEeqnarray}{c}
\dMAT = \frac{K}{1 + \frac{1}{2} + \cdots + \frac{1}{K}} \approx \dfrac{K}{\log(K)},
\label{eq:MAT}
\end{IEEEeqnarray}
i.e. scaling with the number of users $K$. The two main ingredients of their approach are: {\it delayed CSIT precoding} and {\it user scheduling}. Linear combinations (LCs) of all symbols (higher than the number of receive antennas) exploiting the delayed CSIT are sent along all the phases, working similarly to the automatic repeat request (ARQ) protocols, where the same message (or packet) is retransmitted until it can reliably be decoded at the receiver side. Besides, each phase uses a different user scheduling schema, as follows. During phase $p$, the scheme imposes that $K$ users are served in different time instances by groups of $p \leq K$ users, whereas the rest of users listen and learn about the interference, i.e. the unintended messages. This user scheduling, starting from a time division multiplexing access (TDMA) fashion in the first phase, is decided beforehand, independent of the channel realizations, and aims to control the number of interference terms contributing to the signal observed at each receiver. The key idea is to use the obtained LCs of symbols when one user acts as a listening user (thus containing non-intended symbols) as {\it overheard interference} (OHI) that can be aligned (thus removed) by this user, thus it is negligible to it, while being useful for another user. Hence, the delayed CSIT is exploited in such a way that the transmitted signals generate interference at some receivers, but it can be removed thanks to the signals received and buffered in previous phases. This idea allows that more than one user can simultaneously be served after the first phase.

Inspired by the MAT scheme in \cite{MAT}, many works appeared for studying the IC with delayed CSIT, e.g. \cite{Vaze2IC}-\cite{CastanheiraIC}.
To date, only the DoF of the 2-user MIMO IC have been completely characterized, thanks to the work of Vaze et al. in \cite{Vaze2IC}, whereas the case with $K>2$ users is still an open problem. Most of the research has been focused on deriving new DoF inner bounds, i.e. proposing transmission strategies, but there is a lack of DoF outer bounds, with the exception of the work by Lashgari et al. in \cite{rankRatioXC_Trans} for the SISO case. 
In terms of achievable DoF, it is worth pointing out three works: \cite{Hao_MIMOIC}, \cite{TAV_TIT}, and \cite{CastanheiraIC}, each focusing on different aspects. In the former, Hao and Clerckx  proposed a linear precoding scheme (hereafter denoted as the HC scheme) providing the best DoF inner bound for this channel for most antenna settings, but collapsing to a constant value as the number of users grows. In contrast, \cite{TAV_TIT} proposed lower complexity schemes compared to \cite{Hao_MIMOIC}, achieving more competitive \textit{DoF-delay tradeoffs}. In this work, we will also compare the achieved DoF to the number of required time slots to realize them, as most delayed CSIT-based transmission strategies entail prohibitive and not realistic complexity. Finally, the best DoF inner bound for the SISO case (and some other cases with lower number of antennas at receivers compared to  transmitters) was derived in \cite{CastanheiraIC} (hereafter denoted as the CSG scheme, given the authors' surnames), reaching a DoF value \textit{scaling with the number of users}. The main idea introduced in \cite{CastanheiraIC} is the concept of \textit{IA at the receiver side}. However, this is at the cost of assuming an unbounded number of time slots.

As will be explained throughout this paper, the MAT scheme is not directly applicable to the IBC. While some transmission strategies (e.g. null-steering \cite{ZFBD}) for the BC based on linear precoding are transferable to the IC and IBC just by using an almost identical precoding matrix design, in this case it is key to receive all the interference from the same node. Otherwise, the IA is broken. In this regard, in the IBC there are multiple BSs, one per cell, thus not all the interference is received from the same node. We call this phenomena \textit{interference coupling}, as it is not possible to separate the contributions of each BS, required to cancel the interference in posterior phases. Only two works have proposed schemes improving the DoF inner bound for the IBC with completely delayed CSIT: \cite{TAV_IBC_ISIT} and \cite{IBC_dCSIT_ISIT14}, both just for the two-cell case. However, with the exception of a particular setting where the scheme in \cite{TAV_IBC_ISIT} achieves the highest DoF figure ever reported, in most settings all the contributions in both papers are outperformed by applying state-of-the-art originally proposed for other channels. In particular, there are two options. First, using the schemes for the IC, e.g. HC scheme in \cite{Hao_MIMOIC}. And second, using the MAT scheme, together with TDMA so that only one cell is active at a time. This naive approach provides the highest DoF figure for most deployment settings indeed, as the performance of the HC scheme does not scale as previously mentioned.

For sake of completeness, it is worth taking into account the recent work by Huang et al. in \cite{huang2019degrees}, where they propose a \mbox{distributed} space-time interference alignment (DSTIA) approach. Even though the scheme provides remarkable DoF performance assuming local CSIT only, they actually assume a moderate delay, i.e. the terminals have access to delayed CSIT during a fraction $\lambda$ of the coherence time, and obtain the current CSIT afterwards. In this paper we focus on the completely delayed CSIT ($\lambda >1$), being trivially outperformed by simply combining MAT and TDMA. Consequently, \cite{huang2019degrees} will be omitted hereafter.

As reviewed above, the optimal DoF of the IBC with full CSIT are attained by either using or extending ideas originally devised for analogous broadcast or interference channels. Inspired by this concept, we intend to face the following challenges:

\begin{enumerate}
\item Can the performance of the HC, CSG and MAT schemes be improved when applied to the IBC?
\item If so, can the achieved DoF scale with the number of users per cell / number of cells, and with a better scaling factor than MAT?
\item Is all that only possible at the cost of very long transmission delays?
\end{enumerate}

\subsection{Contributions}
\label{sec:contributions}

This work studies the $(L,C)$ MISO IBC with delayed CSIT, where there are $C$ cells, with one BS serving $L$ UEs each. The studied scenario can be understood as having multiple adjacent BCs working at a time, with BSs interfering the other cells UEs. As an example, consider the $(L,2)$ MISO IBC depicted in \mbox{Fig. \ref{fig:scenario}}, with $L$ users per cell and $C=2$ cells. The contributions of this work are summarized next:

\begin{figure}[]
\begin{minipage}[]{1\linewidth}
  \centering 
  \centerline{\includegraphics[width=0.55\linewidth]{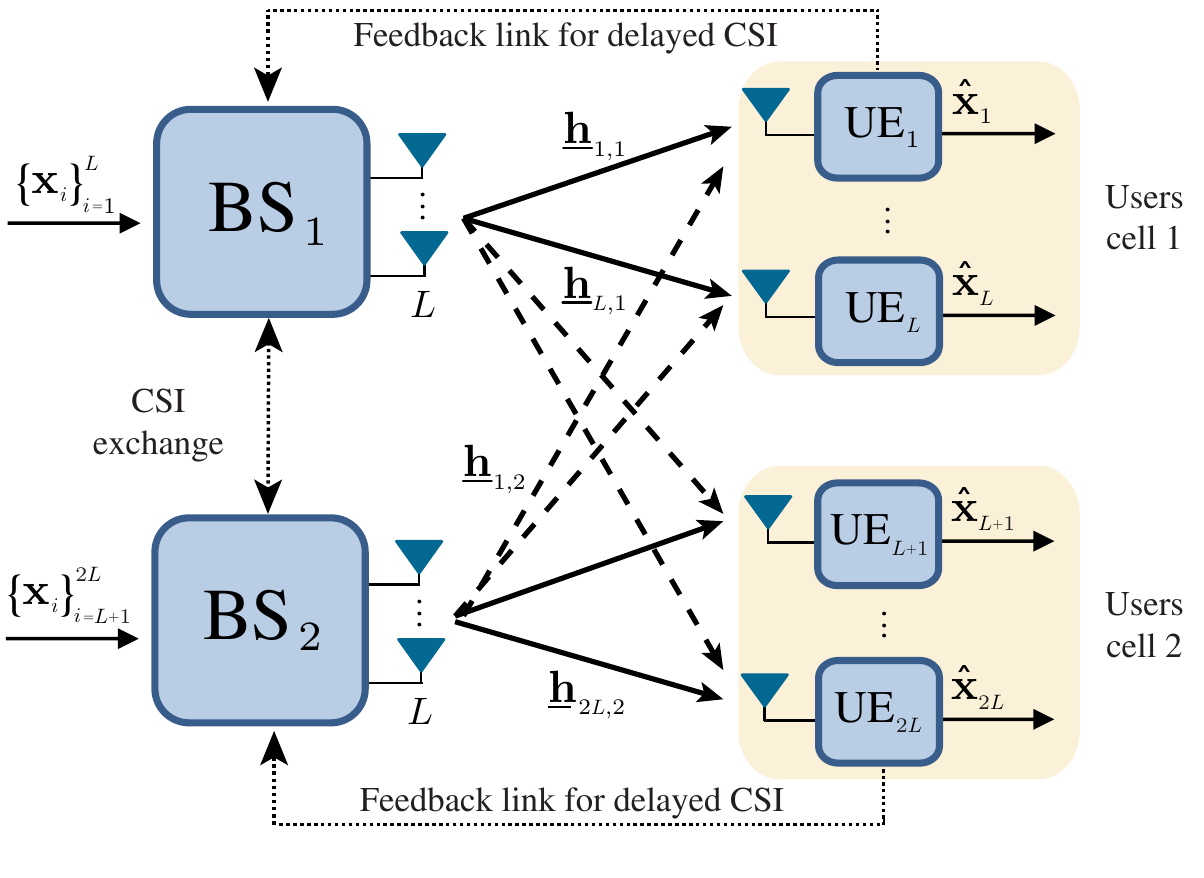}}          
\end{minipage}  
\caption{The $(L,2)$ MISO IBC ($L$ users per cell, 2 cells), with $(L,1)$ antennas at the transmitters and receivers, respectively. Solid lines denote links carrying intended signals, dotted lines denote links for CSI exchange, and dotted lines denote links carrying inter-cell interference signals.}
\label{fig:scenario}
\end{figure}

\begin{itemize}
  \item Inspired on MAT, we propose the uncoupled MAT (uMAT) scheme. While MAT was originally designed for the BC, uMAT allows to \textit{uncouple the interference} from different sources by taking into account the specific topology of the IBC and the concept of \textit{redundancy transmission}.
  
  \item The proposed approach is formulated for general $(L, C)$, obtaining a closed form expression for the achieved DoF.  Except for the case $(C>2, L=3)$\footnote{Actually, the scheme in \cite{Hao_MIMOIC} requires $L^2$ antennas at the BSs for those cases, whereas uMAT only $L$. As usually antennas are more costly at the receiver than the transmitter side, in this paper we assume no limitation on the number of antennas at the BSs.}, the proposed inner bound becomes the best achievable DoF inner bound known for such scenario. Moreover, we prove that the gap between our scheme and previous state-of-the-art schemes monotonically increases as the number of UEs per cell $L$ grows. Based on our experimental results, we conjecture that the same is valid for an arbitrary number of cells $C$.
  
  \item We compute the DoF-delay values for the proposed scheme and compare to previous state-of-the-art, aiming to evaluate the complexity in terms of required number of time slots. Compared to the MAT scheme applied to the IBC, uMAT provides higher DoF at the cost of increased complexity. Compared to the HC scheme, it provides a dramatic increase of DoF at a reduced complexity and number of phases. The CSG scheme is not considered in this case, as it assumes an unbounded number of time slots.
  
 \item While most papers derive the DoF expressions achieved by their proposed schemes, almost none describe how to obtain the number of transmitted symbols per user and time slots for each round and phase to actually implement the scheme. In this regard, we introduce an algorithm that derives the duration of the phases and the number of transmitted symbols for each tuple $(L,C)$. Such algorithm is explained for uMAT, but is almost identical for MAT. A simple extension (omitted to avoid redundancy) has been used to derive the numerical results for the HC scheme.
  
\end{itemize}


\subsection{Organization}

The paper is organized as follows. \mbox{Section \ref{sec:systModel}} introduces the system model considered in this work. Then, \mbox{Sections \ref{sec:MAT} and \ref{sec:HC}} review the MAT and HC schemes, respectively, as we will compare them to uMAT. We also explain how to derive the number of time slots for MAT. Note that because the uMAT scheme builds upon MAT, it is worth understanding its mechanics, and why it fails for the IBC. \mbox{Section \ref{sec:MainResults}} summarizes the main results: DoF inner bounds and DoF-delay benchmarking. The most novel part of this paper is described in \mbox{Section \ref{sec:uMAT}}, where the uMAT scheme is described. Finally, conclusions and future work are drawn in \mbox{Section \ref{sec:conclusion}}.

\subsection{Notation}

Boldface, lower case types refers to column vectors ($\mathbf{x}$). Row vectors are also underlined ($\underline{\mathbf{x}}$), while matrices are written in uppercase types ($\mathbf{X}$). Sets and subspaces are denoted by uppercase types written in calligraphic fonts ($\mathcal{X}$). Furthermore, $\mathbb{C}$ and $\enters$ denote the field of complex numbers, and positive integers, respectively.

We define $\0$ and $\mathbf{I}$ as the all-zero and identity matrices, respectively, with suitable dimensions according to the context. For vectors and matrices, $(\cdot)^T$ is the transpose operator, and $(\cdot)^H$ is the transpose and conjugate operator. Moreover, the following two predefined vector and matrix operations are defined:
\begin{IEEEeqnarray*}{c c c}
\def\arraystretch{1.2}
\stack{\Xn,\Yn} =
\begin{bmatrix}
\Xn \\
\Yn \\
\end{bmatrix},
& \quad &
\bdiag{\Xn,\Yn} =
\begin{bmatrix}
	\Xn & \0 \\
	\0 & \Yn \\
\end{bmatrix}.
\end{IEEEeqnarray*}

$\Span{\Xn}$ is usually used to define the \textit{column subspace}, containing all possible linear combinations of the columns. However, in this work we always use the \textit{row subspace}, defined as ${\Xc = \RSpan{\Xn}=
\mathsf{span} \big( \Xn^T \big) }$, whose dimension is given by 
$\Dim{\Xc}={\Rank{\Xn}}$. 

Given a fraction $\frac{a}{b}$, we denote the members of its irreducible version as $\bar{a}, \bar{b}$, i.e. $\frac{a}{b} = \frac{\bar{a}}{\bar{b}}$, which can be obtained as 
\begin{IEEEeqnarray}{c}
\bar{a} = \frac{a}{\mathsf{gcd}\left(a,b\right)}, \quad \bar{b} = \frac{b}{\mathsf{gcd}\left(a,b\right)},
\label{eq:gcd}
\end{IEEEeqnarray}
where $\mathsf{gcd}(a, b)$ denotes the greatest common divisor of $a$ and $b$. We also define $\mathsf{lcm}(a, b)$ as the least common multiplier of $a$ and $b$. Finally, we assume that a summation for 0-length set equals 0, whereas a product for a 0-length set equals 1.

\section{System model}
\label{sec:systModel}

In this section we introduce more formally the scenario considered for this work, along with a general transmission protocol that can be used to describe any of the schemes considered in this work. We also review some of the general concepts in the literature that will be used throughout the next sections, namely delayed CSIT constraint, time sharing, and high-order DoF.

\subsection{Preliminaries}

This works addresses the $(L,C)$ MISO IBC, where $C$ transmitters serve (sharing the same frequency band) $L$ users each, see \mbox{Fig. \ref{fig:scenario}}. The total number of users is given by
\begin{IEEEeqnarray}{c}
K = L \cdot  C.
\end{IEEEeqnarray}
While we do not assume a limitation on the number of transmitter antennas at the BSs, unless otherwise stated we will refer to BSs equipped with $L$ antennas. Each $\BS{i}$ aims to deliver $b$ independent symbols to each single-antenna receiver $\UE{j}$ in its cell, i.e. the users in the set:
\begin{IEEEeqnarray}{c}
\mathcal{C}_i = \left\{ j \in \{ 1, \dots, K \} \, | \, c(j) = i \right\}.
\label{eq:cell_users}
\end{IEEEeqnarray}
with 
\begin{IEEEeqnarray}{c}
 c(j)= \Big\lceil \frac{j}{L} \Big\rceil \in \{ 1, \dots ,C \}.
\end{IEEEeqnarray}

\begin{figure}[]
\begin{minipage}[]{1\linewidth}
  \centering  
   \centerline{\includegraphics[width=0.45\linewidth]{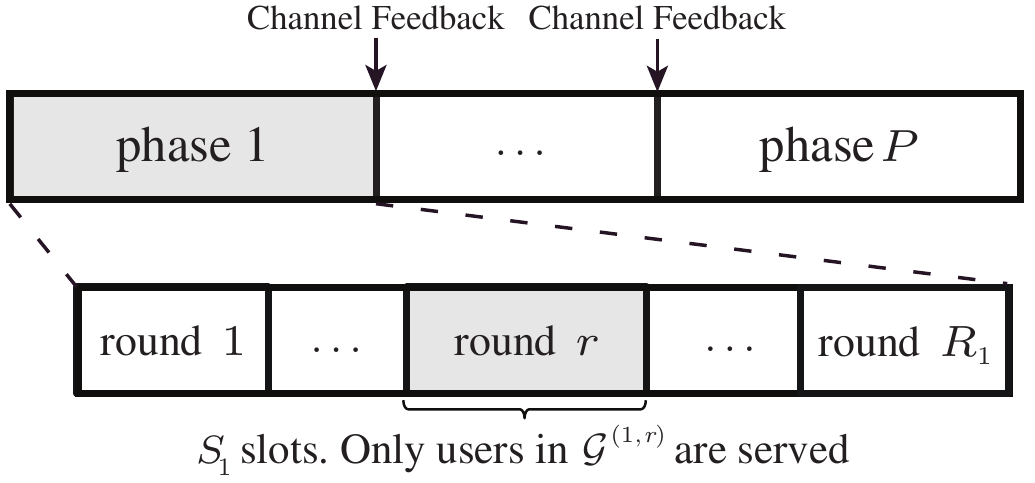}}
\end{minipage}    
\caption{General protocol frame for all the schemes described in this paper. After each phase, CSI is obtained at the transmitters through feedback. There are $P$ phases, where the phase $p$ is divided into $R_p$ rounds. The group of $p$ users served during the $(p,r)th$ round (\textit{served users}) is predefined by the set $\Ac^{(p,r)}$. Finally, each round of the phase $p$ is in turn divided into $S_p$ time slots.}
\label{fig:frame}  
\end{figure}

Communication is carried out in $P$ phases, with each phase $p$ in turn divided into $R_p$ rounds of duration $S_p$ time slots each, see \mbox{Fig. \ref{fig:frame}}. The total number of slots used for data transmission are
\begin{IEEEeqnarray}{c}
\tau= \sum_{p=1}^P \tau_p \,, \quad \tau_p = R_p S_p.
\label{eq:tau}
\end{IEEEeqnarray}
During the $(p,r)$th round, i.e. round $r$ of phase $p$, only a specific group of $p$ users is served, denoted by $\Ac^{(p,r)}$. The users served during a round are denoted as the \textit{served users}, whereas the rest of users will be referred to as the \textit{listening users}.

\subsection{Signal Model}

The signal received at $\UE{j}$ during all the slots of round $(p,r)$ writes as
\begin{IEEEeqnarray}{c}
\yn_j^{\left(p,r\right)} = \Hns{j,c(j)}{p,r} \Vns{j}{p,r} \mathbf{x}_j + 
   \sum_{i = 1}^C  
    \Hns{j,i}{p,r} \!\!\!\!
    \sum_{\substack{k \in \mathcal{C}_i , k \neq j  }} \!\!\!\!\! \Vns{k}{p,r} \mathbf{x}_k + 
    \mathbf{n}_j^{(p,r)},
\label{eq:IBCmodel}
\end{IEEEeqnarray}
where $\yn_j^{\left(p,r\right)} \Cmat{S_p}{1}$ is the vector containing the signals observed at $\UE{j}$, $\mathbf{x}_j \Cmat{b}{1}$ contains the $b$ uncorrelated unit-powered complex-valued data symbols intended to $\UE{j}$. Note that linear combinations of the \textit{same} $b$ symbols are transmitted during all phases, but receivers will not be able to linearly decode them until the end of the communication either because the reduced number of receive antennas, or because of interference\footnote{This paper focuses on linear precoding and receiving schemes. While other approaches treat interference at the signal level (see e.g. \cite{GouJafarImperfect}\cite{Mohanty_GDOF_IC}), here we manage interference at the spatial domain.}. Also, it is worth pointing out that in contrast to the IC, for the IBC all $\mathbf{x}_k, k \in \mathcal{C}_i$ are transmitted by the same $\BS{i}$, thus they are observed through the same channel at each receiver.

Besides, $\Vn_j^{\left(p,r\right)}\Cmat{LS_p}{b}$ denotes the precoding matrix carrying the signals desired by $\UE{j}$, designed subject to a maximum transmission power per user\footnote{This criteria could also be w.r.t. maximum transmission power at each BS. In either case, since we are analyzing the DoF, an unbounded transmission power is assumed.} $\gamma$, and with ${\Vn_j^{\left(p,r\right)}=\0,\forall i \notin \Ac^{(p,r)}}$; and $\mathbf{n}_j^{(p,r)} \Cmat{S_p}{1}$ is the unit-powered noise term, thus the maximum transmission power $\gamma$ denotes also the SNR. It is worth pointing out that the noise terms will be omitted hereafter, as this work is focused on analyzing the DoF, thus assuming arbitrarily high SNR.

The channel coefficients for each slot and each link transmitter-receiver are described by a $1\times L$ row vector. Then, the channel matrix $\Hn_{j,i}^{\left(p,r\right)} \Cmat{S_p}{LS_p}$ in (\ref{eq:IBCmodel}) is formed as the block-diagonal composition of $S_p$ of such vectors, thus contains the channel gains from antennas of $\Tx{i}$ to $\Rx{j}$ during all time slots of the $(p,r)$th round. For simplicity, it is assumed that channel coefficients follow a flat block fading channel model, i.e. they are i.i.d. as $\mathcal{CN}\!\pare{0,1}$, and completely uncorrelated in time and space\footnote{While this is the usual assumption across all delayed CSIT literature, some schemes can also be used in case transmitters assume delayed CSIT, but channels are actually constant. This case was studied in \cite{TAV_TIT} for the IC.}.

The objective of the receivers is retrieving $b$ linear combinations of its desired symbols from the received signals. To do so, they collect the signals along the whole communication, so that the per-round expression in (\ref{eq:IBCmodel}) can be written in compact form as \cite{TAV_TIT}
\begin{IEEEeqnarray}{c}
\label{eq:SystemModelextended}
\def\arraystretch{1.2}
\begin{matrix}
\yn_j = \stack{\yn_j^{(1)},\dots,\yn_j^{(P)}} =  \Gn_j \cdot \stack{\xn_1, \dots,\xn_K }, 
\end{matrix}
\end{IEEEeqnarray}
with
\begin{IEEEeqnarray}{c}
\label{eq:SystemModelextended2}
\begin{matrix}
\yn_j^{(p)} = \stack{\yn_j^{(p,1)},\dots,\yn_j^{(p,R_p)}} =  \Gn_j^{(p)} \cdot \stack{\xn_1, \dots,\xn_K }, \\[2mm]
	\Gn_j = \stack{\Gn_j^{(p)},\dots,\Gn_j^{(p)}} = 
    \begin{bmatrix} 
		\Hn_{j,1} \Vn_{1}, \, \ldots ,\,
		\Hn_{j,K} \Vn_{K}
	\end{bmatrix}, \\[2mm]
\Hn_{j,i} = \bdiag{
\begin{matrix}
	 \Hn_{j,i}^{\left(1\right)}, \dots,
	\Hn_{j,i}^{\left(P\right)}
\end{matrix} } , \quad
\Vn_{i} = \stack{
\begin{matrix}
	\Vn_{i}^{\left(1\right)}, \dots,
	\Vn_{i}^{\left(P\right)}
\end{matrix} }, \\[2mm]
\Hn_{j,i}^{(p)} = \bdiag{
\begin{matrix}
	\Hn_{j,i}^{\left(p,1\right)}, \,
	\Hn_{j,i}^{\left(p,2\right)}, \,\ldots, \,
	\Hn_{j,i}^{\left(p,R_p\right)}
\end{matrix} }, \\[2mm]
\Vn_{i}^{(p)} = \stack{
\begin{matrix}
	\Vn_{i}^{\left(p,1\right)}, \,
	\Vn_{i}^{\left(p,2\right)}, \,\ldots, \,
	\Vn_{i}^{\left(p,R_p\right)}
\end{matrix} },
\end{matrix}
\end{IEEEeqnarray}
\\[-2mm]
where $\Gn_j \Cmat{\tau}{Kb}$ is the Signal Space Matrix (SSM) \cite{TAV_ppM1}, defining the subspaces occupied by the received signals at each receiver, and
${\Hn_{j,i} \Cmat{\tau}{L\tau}}$, ${\Vn_i\Cmat{L\tau}{b}}$, 
$\Gn_j^{(p)} \Cmat{\tau_p}{Kb}$, $\yn_j^{(p)} \Cmat{\tau_p}{1}$, ${\Hn_{j,i}^{(p)} \! \Cmat{\tau_p}{L\tau_p}}$, and ${\Vn_{i}^{(p)} \Cmat{L\tau_p}{b}}$. 

\subsection{Delayed CSIT Constraint}

All precoding and receiving filters are designed subject to a delayed CSIT model.  This means all CSI is instantaneously assumed to be known at the receiver side, whereas only the channels
\begin{IEEEeqnarray*}{c}
 \{ \Hns{j,i}{\varrho}\}_{\varrho=1}^{p-1},\forall i,j,
\end{IEEEeqnarray*}
are available at the transmitter side at the beginning of phase $p$. In other words, each transmitter knows all past CSI for all links and phases, and because channels are i.i.d., it has no knowledge about current CSI.

\subsection{Degrees of Freedom}

The channel DoF are defined as the slope of the maximum achievable bit-rate (\textit{channel capacity}) $B(\gamma)$ for arbitrarily high SNR $\gamma$ \cite{Cover}:
\begin{IEEEeqnarray}{c}
d \! = \lim_{\gamma \to \infty} \,\, \frac{B(\gamma)}{\log_2 \gamma}.
\label{eq:DoFdef}
\end{IEEEeqnarray} 
A typical method to characterize the DoF is by proving that one inner bound and one upper bound match. A DoF inner bound might be obtained by computing the achievable bit-rate of a given transmission strategy and applying (\ref{eq:DoFdef}). However, many works assume that simple receiving filters ${\Wn_j \Cmat{b}{\tau}}$ are used, with 
\begin{IEEEeqnarray}{c}
\mathbf{{W}}_{j} \mathbf{{H}}_{j,c(k)} \mathbf{{V}}_{k}  = \0 , \quad \forall k \neq j ,
\label{eq:noInterferenceb}
\end{IEEEeqnarray} 
i.e. acting as a linear zero-forcing filter that projects the received signals onto the orthogonal-to-interference space, thus separating desired signals from interference. In such a case, the achievable DoF write as (see \cite{TAV_ppM1},\cite{TAV_TIT})
\begin{IEEEeqnarray}{c}
\label{eq:achDoF}
\din = 
\frac{1}{\tau} \sum_{j = 1}^K \Rank{ \Wn_j \, \Gdes  }  
\overset{(a)}{\leq}
\frac{Kb}{\tau},
\end{IEEEeqnarray}
where $\Gdes = \Hn_{j,c(j)} \Vn_j$ collects the columns of the SSM related to the desired signals, and inequality $(a)$ is satisfied with equality only if, after projection, the equivalent channel for each user has rank $b$. In other words, after projection each receiver should be able to retrieve $b$ independent and free of interference LCs or observations of its desired symbols. While in some cases a rigorous proof is required (see \cite{TAV_ppM1}), usually inequality $(a)$ in (\ref{eq:achDoF}) is automatically satisfied with equality with probability one. This is because the precoding matrices are designed to manage the interference, thus direct channels do not take part on the precoding matrix design, and the product of the two independently designed matrices $\Wn_j$ and $\Gdes$ is full-rank.

Finally, notice that we use superscripts to denote what the given DoF count accounts for, e.g. $\din$ stands for DoF inner bound.

\subsection{High-Order Symbols and DoF}

The MAT scheme \cite{MAT} introduced the concept of \textit{order-$p$ symbols}, as symbols simultaneously useful or intended to $p$ users. Specifically, an order-$p$ symbol desired by the users in the set $\Tc$ is a linear combination of all the symbols in the set $\{ \xn_i \, | \, i \in \Tc \}$. For example, consider $p=2$, and $\Tc = \{ i,j \}$, then
\begin{IEEEeqnarray}{c}
\label{eq:ordermsymbol}
\un(\Tc) =\un(i,j) =  \boldsymbol{\Sigma} \bigg[ \, \begin{matrix} \xn_i \\[-2.5mm] \xn_j \end{matrix}\, \bigg]
\end{IEEEeqnarray}
denotes an order-2 symbol intended to the two users in $\Tc$, where $\boldsymbol{\Sigma}$ is a matrix that linearly combines the users' symbols, and depends on each case.

The efficiency on transmitting order-$p$ symbols (i.e. symbols that are intended to a group of $p$ users) is measured by the \textit{order-$p$ degrees of freedom}, and denoted hereafter by $d_p$. Notice that the conventional DoF may be found by definition as
\begin{IEEEeqnarray}{c}
\label{eq:d1}
d  \triangleq d_1.
\end{IEEEeqnarray} 

Interestingly, we will see later that most works (see e.g. \cite{MAT}) formulate order-$p$ DoF as a function of the order-($p+1$) DoF, e.g:
\begin{IEEEeqnarray}{c}
\label{eq:dmm1}
d_p = f \left( d_{p+1} \right),
\label{eq:recursiveDoF}
\end{IEEEeqnarray}
for some function $f$, enabling generalization to any $p$ based on the induction method.

\subsection{Time-Sharing}
\label{sec:timesharing}

Any scheme working for $Q<C$ cells can be used for the $(L,C)$ IBC by subsequently activating only $Q$ cells and turning off the BSs and UEs located in the $(C-Q)$ non-active cells. Then, after all, all possibles groups of $Q$ cells are served once. 
When this approach is applied, the same DoF sum achieved in the $(L,Q)$ IBC is achieved for the $(L,C)$ case, but the number of time slots $\tau$ is multiplied by $\binom{C}{Q}$.

\section{Review of the MAT Scheme}
\label{sec:MAT}

The IBC with a single cell ($C=1$) reduces to the $K$-user BC. This section reviews the MAT scheme  \cite{MAT}, able to attain the optimal DoF for that channel. First, we review the main theorem, and give the intuition of the proof, deferring the details to the full paper. Second, we explain in detail a truncated example for $K=6$ users. Finally, the MAT scheme is shown to fail when applied to the IBC, encouraging the efforts towards deriving improved schemes like the one proposed in this work.

\subsection{Main Results}
\label{sec:MAT:main}

The optimal scheme for the BC consists of $P=K$ phases. During each phase $p$, order-$p$ symbols are transmitted, with an efficiency $\dMAT_p$, and formulated in a recursive way as in (\ref{eq:recursiveDoF}). The main result in \cite{MAT} is next stated:
\\
 
\begin{theorem}[MAT scheme, achievable order-$p$ DoF \cite{MAT}]
For the $K$-user MISO BC with delayed CSIT and $K$ antennas at the transmitter, assuming that $\dMAT_{p+1}$ DoF can be achieved, then
 \label{th:MATrec} 
\end{theorem}
\begin{IEEEeqnarray}{c}
 \dMAT_p = \frac{(K-p+1) \binom{K}{p}} 
	    { \binom{K}{p} + \frac{p \binom{K}{p+1}} {\dMAT_{p+1} }}
\label{eq:MAT:DoFp}
\end{IEEEeqnarray}
DoF are achievable.\\

In other words, if there exists a transmission strategy allowing to deliver order-$(p+1)$ symbols at $\dMAT_{p+1}$ DoF, then they show how to achieve the order-$p$ DoF value in (\ref{eq:MAT:DoFp}). While the full proof may be found in \cite{MAT}, here the methodology and concepts behind the MAT scheme are briefly summarized. Details for a truncated example may be found in the next section.

Each round of phase $p$ there are exactly $p$ served users. The total number of rounds for such phase is given by:
\begin{IEEEeqnarray}{c}
R_p^{(\text{MAT})} = \binom{K}{p},
\label{eq:MAT:Nrounds}
\end{IEEEeqnarray}
i.e. all possible groups of $p$ users out of $K$. During the $p$th phase, $(K-p+1)$ symbols of order $p$ are transmitted per round. As the DoF can be seen as the ratio between the number of symbols and the number of time slots used to transmit them, the arguments explained so far justify both the numerator and first term of the denominator in (\ref{eq:MAT:DoFp}).

Now, the aim is to justify that those symbols can be decoded by using
\begin{IEEEeqnarray}{c}
\frac{p \binom{K}{p+1}} {\dMAT_{p+1} }
\label{eq:MAT:den2}
\end{IEEEeqnarray}
time slots. To this end, consider the following three ideas for the $r$th round of the $p$th phase:
\begin{itemize}
\setlength\itemsep{0.3em}
\item The $(K-p+1)$ transmitted symbols are desired only by the $p$ users (served users) in $\Ac^{(p,r)}$, who obtain one LC of those symbols each.
\item The other $(K-p)$ users (listening users) obtain undesired information, i.e. the received LC do not contain its intended symbols.
\item If each user in $\Ac^{(p,r)}$ obtained the $(K-p)$ LCs received during this phase at the listening users, together with the observation obtained, it would be able to decode the $(K-p+1)$ order-$p$ transmitted symbols.
\end{itemize}
These ideas apply to each round of the $p$th phase. Now, the order-$(p+1)$ symbols for the next phase are constructed as follows. Consider any group of served users for the next phase, e.g. $\Ac^{(p+1,r)}$. For each subset $\Tc$ of $p$ users, there exists an observation available at the remaining user, and desired by all users in $\Tc$. Then, a LC of those $p$ observations (there are $p$ possible subsets $\Tc$ of $p$ users in $\Ac^{(p+1,r)}$) would contain information desired by all $p+1$ users in $\Ac^{(p+1,r)}$. Consequently, $p$ observations are transmitted to each possible group of $p+1$ users, i.e. $\binom{K}{p+1}$, with efficiency $\dMAT_{p+1}$. The corresponding number of time slots are written as (\ref{eq:MAT:den2}), and the proof is complete.

Based on \mbox{Theorem \ref{th:MATrec}}, fixing $\dMAT_K = 1$, and solving the recursive equation in (\ref{eq:MAT:DoFp}), the DoF in (\ref{eq:MAT}) were found, here repeated for reader's convenience:
\begin{IEEEeqnarray}{c}
\dMAT = \frac{K}{1 + \frac{1}{2} + \cdots + \frac{1}{K}} \approx \dfrac{K}{\log(K)},
\label{eq:MATbis}
\end{IEEEeqnarray}

\subsection{Truncated example}

While the MAT scheme requires $K$ phases to attain the optimal DoF for the BC, it is possible to truncate the scheme to $\Theta$ phases\footnote{Notice that in the extreme case $\Theta = 1$, the MAT scheme reduces to TDMA, with $d_{\text{TDMA}} = 1$.} simply by setting $d_{\Theta} =1$ for some $\Theta < K$, at the cost of reducing the achieved DoF. 
As an example, we describe the MAT scheme for $K=6$ users, truncated to $P=3$ phases, thus setting $\Theta = 3 < 6$. To avoid formulation conflicts, we will refer to this scheme as TMAT.

The aim of this section is threefold. First, it allows us to describe the MAT scheme up to the third phase, but with six users, so giving a good overview of the methodology for the design of precoding matrices with a concrete example. We believe that this will help the reader to familiarize with notation and formulation. Second, TMAT is one step further on the way from evolving MAT into uMAT. And finally, this section will be necessary to support Section {\ref{sec:tMAT:IBC}}, where the limitations of MAT and TMAT when applied to the IBC are showcased.

This section will be divided into four parts. First, the parameters $b^{\text{(TMAT)}}$, $R_p^{\text{(TMAT)}}$, $S_p^{\text{(TMAT)}}$ are derived for each round and phase. Then, the transmission strategy is described for each phase.
\\

\subsubsection{Derivation of Parameters}

Starting from $\dTMAT_3 = 1$, and using \mbox{Theorem \ref{th:MATrec}}:
\begin{IEEEeqnarray}{c}
 \dTMAT_2 = \frac{5 \binom{6}{2}} 
	    { \binom{6}{2} + 2 \binom{6}{3}}.
\label{eq:dTMAT2}
\end{IEEEeqnarray}
Then, the achievable DoF are given by:
\begin{IEEEeqnarray}{c}
 \dTMAT = 
	    6 \frac{6 \cdot 5}
	    {6 \cdot 5 + 15 \cdot 1 + 20 \cdot 2} = \frac{36}{17},
\label{eq:dTMAT}
\end{IEEEeqnarray}
where the number of rounds per phase, and number of slots for each round of each phase can easily be identified, see \mbox{Table \ref{tab:paramsTMAT}}. Similarly, from (\ref{eq:dTMAT}) we identify that $b^{\text{(TMAT)}} = 30$ and $\tau^{\text{(TMAT)}}= 85$.

\begin{table}[h]
\caption{System parameters for the 6-user MAT scheme truncated to 3 phases}
\begin{center}
\begin{tabular}{c | c | c | c  | c }
Phase $(p)$ & I & II & III  & Total\\  \cline{1-5} 
Rounds $(R_p^{\text{(TMAT)}})$ & 6 & 15 & 20 & \\  \cline{1-4} 
Slots per round $(S_p^{\text{(TMAT)}})$ & 5 & 1 & 2 & \\ \cline{1-5}
Slots per phase $(\tau_p^{\text{(TMAT)}})$ & 30 & 15 & 40 & 85 \\ \cline{1-5} 
 LCs acquired per  & \multirow{2}{*}{5} & \multirow{2}{*}{5} & \multirow{2}{*}{20}  &  \multirow{2}{*}{30}\\[-2mm]
phase and user &  &  & 
\end{tabular} 
\end{center}
\label{tab:paramsTMAT}
\end{table} 

\subsubsection{Phase I}

The first phase consists of six rounds of five slots each. During each round, only one user is served per round, i.e: $\Ac^{(1,r)} = r$. 
The precoding matrices used during this round are known by all terminals and assigned randomly from a codebook, known prior to the communication by all parties involved. Consequently, after the first phase each receiver acquires $S_1^{\text{(TMAT)}} = 5$ LCs of desired symbols, thus $b^{\text{(TMAT)}} - S_1^{\text{(TMAT)}} = 25$ additional LCs are required. Interestingly, those remaining LCs are distributed along the listening receivers, who observe 5 LCs of each non-desired set of symbols. These LCs will be the basis to align the interference during the next phases, while providing new LCs of desired symbols. They represent the \textit{overheard interference} for the first phase, denoted by  
\begin{subequations} 
\begin{IEEEeqnarray}{c}
 {\Tns{j,i}=\Hns{j}{1,i}   \Vn_{i}^{\left(1,i\right)} \Cmat{5}{30}}, \\[2mm]
 \Tc_{i,j} = \RSpan{\Tns{j,i}},
\end{IEEEeqnarray}
\end{subequations} 
i.e. the signals intended to $\UE{i}$ and observed at $\UE{j}$. Moreover, notice that all these $30$ distributed LCs can be assumed linearly independent with probability one due to the channel randomness. Finally, for better readability we show the signal space matrix for $\UE{j}$ after the first phase:
\vspace{-4mm} \\
\begin{IEEEeqnarray}{c}
\def\arraystretch{1.4}
\Gn_j^{(1)} = \bdiag{ \Tns{j,1},\Tns{j,2},\ldots,\Tns{j,6}} = 
\left[ \,
\begin{matrix}
   \Tns{j,1} & \0 &     \cdots   & \0       \\ 
      \0  &  \Tns{j,2}  &     \cdots   & \0       \\ 
         \vdots & \vdots &  \ddots  & \vdots       \\ 
            \0 & \0 & \cdots &  \Tns{j,6}  
\end{matrix} \, \right].
\label{eq:TMATG1}
\end{IEEEeqnarray}

\begin{table}[]
\caption{Served users per round, and acquired OHI at $\UE{1}$ for each round of the 6-user MAT scheme truncated to 3 phases.\hspace{\textwidth} Only exploited OHI is specified.}
 \begin{center} 
 \def\arraystretch{2}
 \begin{tabular}{C | C | C  | C}
 \hline
 \text{Phase} \, (p) \!\!\!& \! \text{Round} \, (r) \!\!\!& \Ac^{(p,r)} & \text{Acquired OHI at $\UE{1}$} \\
 \hline
 \multirow{4}{*}{1} & 1 & \{1\} &  - \\
  & 2 & \{2\} &  \Tns{1,2} \, \xn_2 \\
  & \vdots & \vdots &  \vdots \\
  & 6 & \{6\} &  \Tns{1,6} \, \xn_6 \\
  \hline
 \multirow{6}{*}{2} & 1 & \{1,2\} &  - \\
 & 2 &  \{1,3\} &  - \\
 & \vdots &  \vdots &  \vdots \\
 & 14 &  \{4,6\} &  \tng{6,4}{1} \, \xn_4 + \tng{4,6}{1} \, \xn_6 \\
 & 15 &  \{5,6\} &  \tng{6,5}{1} \, \xn_5 + \tng{5,6}{1} \, \xn_6 \\
 \hline
 \multirow{4}{*}{3} & 1 & \{1,2,3\} & - \\
 & 2 &  \{1,2,4\} & - \\
 & \vdots &  \vdots &  \vdots \\
 & 20 &  \{4,5,6\} &  - \\
 \end{tabular} 
 \end{center} 
 \vspace{-5mm}
\label{tab:OHI_TMAT}
\end{table}

\subsubsection{Phase II}

Order-2 symbols are transmitted during this phase, thus for each round, pairs of users are served. Specifically, there are 15 rounds of a single slot each. In this case, the first phase OHI is exploited to design the transmitted signals. Consider the round $r$, dedicated to $\UE{i}$ and $\UE{j}$, i.e. $\Ac^{(2,r)}\!\!=\!\{i,j\}$. According to the procedure described in \mbox{Section \ref{sec:MAT:main}}, the following order-2 symbol is transmitted:
\begin{IEEEeqnarray}{c}
 \un (i,j)  = \Sigmas{i}{2,r} \Tns{j,i} \, \xn_i +  \Sigmas{j}{2,r} \Tns{i,j}  \, \xn_j		 \label{eq:HOS2},
\end{IEEEeqnarray}
where the precoding matrices are identified as
\begin{IEEEeqnarray}{c}
 \Vns{i}{2,r}  = \Sigmas{i}{2,r} \Tns{j,i} \label{eq:scC:desPhase2}, \quad
 \Vns{j}{2,r}  = \Sigmas{j}{2,r} \Tns{i,j},
 \label{eq:precPhase2}
\end{IEEEeqnarray}
and $\Sigmas{i}{2,r}, \Sigmas{j}{2,r}  \Cmat{6}{5} $ are some random full rank matrices ensuring the transmit power constraint. 
\\[1.5mm]
Now consider the signal received at user $\UE{k}$, $k \notin \{ i, j \}$, during this round:
\begin{IEEEeqnarray}{c}
 \yn_k^{(2,r)} = \tng{j,i}{k} \xn_i + \tng{i,j}{k} \xn_j,
\label{eq:OHIx2}
\end{IEEEeqnarray}
where
\begin{subequations} 
\begin{IEEEeqnarray}{c}
 \tng{j,i}{k} = \hns{k}{2,r} \Vns{i}{2,r}   \Cmat{1}{30}, \\
  \tng{i,j}{k} = \hns{k}{2,r} \Vns{j}{2,r}  \Cmat{1}{30}.
\end{IEEEeqnarray}
\label{eq:OHI2}
\end{subequations}
Notice that $\tng{j,i}{k}$ is the LC of $\Tns{j,i}$ observed at $\UE{k}$, in general different from the one observed at other receivers. As an example, consider the 10th round of this phase, where $\Ac^{(2,10)} = \left\{ 3,4 \right\}$. Then, the term
$ {\tng{4,3}{5} = \hns{5}{2,10} \Sigmas{3}{2,10} \Tns{4,3}}$ represents the LC of 
$\Tns{4,3}$ observed at $\UE{5}$ (so $k = 5$) during that round, and expanding (increasing the dimension of) its interference subspace. However, as:
\begin{IEEEeqnarray}{c}
        \RSpan{ \tng{j,i}{j} } 
        \subseteq 
        \RSpan{ \Tn_{j,i}} = \Tc_{j,i}, \label{eq:RIA_phase2b} 
\label{eq:RIA_phase2b}
\end{IEEEeqnarray}
the part of the signal received at $\Rx{j}$ during this round corresponding to interference is already known from the first phase OHI, which is typically referred as being \textit{aligned} with the OHI gathered during the first phase. This means that one can use such buffered received signals to cancel the interference, and effectively each user in $\Ac^{(2,r)}$ obtains one extra LC of desired symbols.

Overall, the second phase SSM at $\UE{j}$ writes as follows:
\begin{IEEEeqnarray}{c}
\def\arraystretch{1.6}
\Gn_j^{(2)} = \left[ \,
\begin{matrix}
   \tng{2,1}{j} & \tng{1,2}{j} & \0 & \0 & \0 & \0 \\
   \tng{3,1}{j} & \0 & \tng{1,3}{j}  & \0 & \0 & \0\\
   \vdots & \vdots & \vdots & \vdots & \vdots & \vdots \\
   \0 & \0 & \0  & \tng{6,4}{j} & \0 & \tng{4,6}{j} \\
   \0 & \0 & \0  & \0 & \tng{6,5}{j} & \tng{5,6}{j} 
\end{matrix} \, \right].
\label{eq:TMATG2}
\end{IEEEeqnarray}

Finally, from the received signals represented by (\ref{eq:TMATG1}) and (\ref{eq:TMATG2}), each user is able to retrieve 5 new LCs of desired symbols by combining the current received signals and previous OHI. Consequently, 20 LCs of desired signals are still required to decode all the $b=30$ desired symbols.

\subsubsection{Phase III}

Triplets of users are served during each of the $R_3 = 20$ rounds of this phase, consisting of $S_3 = 2$ slots each. 
Similarly to the second phase, the second phase OHI is exploited to design the transmitted signals taking into account the following two premises:
\begin{itemize}
\item Desired symbols are delivered to each of the three users in $\Ac^{(3,r)}$.
\item The interference listened at the served users can be removed thanks to interference alignment.
\end{itemize}

Consider the round $r$ dedicated to $\UE{i}$, $\UE{j}$, and $\UE{k}$, i.e. $\Ac^{(3,r)}\!\!=\!\{i,j,k\}$. Then, the following order-3 symbol is transmitted during the two times slots of this round:
\begin{subequations} 
\begin{IEEEeqnarray}{r l}
 \un (i,j,k)  \, \, &  = \left[ \sigmas{1}{3,r} ,  \sigmas{2}{3,r} \right]
 			\begin{bmatrix} 
 				\tng{j,i}{k} \\
				\tng{k,i}{j} 
			  \end{bmatrix}	
			\, \xn_i 
		+  \left[ \sigmas{1}{3,r} ,  \sigmas{3}{3,r} \right]
 			\begin{bmatrix} 
 				\tng{i,j}{k}  \\
				\tng{k,j}{i} 
			  \end{bmatrix}	
			\, \xn_j
		+  \left[ \sigmas{2}{3,r} ,  \sigmas{3}{3,r} \right]
 			\begin{bmatrix} 
 				\tng{i,k}{j}  \\
				\tng{j,k}{i} 
			  \end{bmatrix}	
			\, \xn_k 
					 \label{eq:HOS3_1}
			\\[2mm]
& = \sigmas{1}{3,r} \left( \tng{j,i}{k} \xn_i+ \tng{i,j}{k} \xn_j \right) +
   \sigmas{2}{3,r} \left( \tng{k,i}{j} \xn_i+ \tng{i,k}{j} \xn_k \right) +
   \sigmas{3}{3,r} \left( \tng{k,j}{i} \xn_j+ \tng{j,k}{i} \xn_k \right),
		 \label{eq:HOS3_2}
 \label{eq:order2_TMAT}
\end{IEEEeqnarray}
\end{subequations} 
where $\sigmas{l}{3,r} \Cmat{12}{1}, \forall l \in \Ac^{(3,r)}, $ are some random vectors ensuring the transmit power constraint. As before, notice that from (\ref{eq:HOS3_1}) the precoding matrices  $ {\Vns{l}{3,r} \Cmat{12}{30} }$ can easily be identified. Notice that in (\ref{eq:order2_TMAT}) we rewrite the expression to give emphasis to the fact that the OHI observed at the listening UEs during the second phase (see (\ref{eq:OHIx2})) is being retransmitted.

Based on this design, all the interference is aligned and thus can be removed at all receivers in $\Ac^{(3,r)}$. For example, consider $\UE{i}$, where the following properties hold:
\begin{subequations}
\begin{IEEEeqnarray}{c}
\RSpan{ 
        \Hns{i}{3,r} \sigmas{1}{3,r} \tng{i,j}{k} } 
       \subseteq 
                \Tc_{i,j}, \label{eq:TMAT_RIA_31} \\
\RSpan{ 
        \Hns{i}{3,r} \sigmas{2}{3,r} \tng{i,k}{j} } 
       \subseteq 
               \Tc_{i,k}, \label{eq:TMAT_RIA_32} \\
\RSpan{ 
        \Hns{i}{3,r} \sigmas{3}{3,r} \left[ \tng{k,j}{i} ,  \tng{j,k}{i} \right] } 
       \subseteq 
               \RSpan{ \left[ \tng{k,j}{i} ,  \tng{j,k}{i} \right]}. \label{eq:TMAT_RIA_33}
\end{IEEEeqnarray}
\label{eq:TMAT_RIA3}
\end{subequations} 

After removing the interference, each round provides two LCs of desired symbols, one for each slot, to each served user. Consequently, since each user appears in 10 out of the 20 possible groups of three users, after this phase the remaining 20 LCs are obtained, and finally all symbols can linearly be decoded.

\subsection{Applicability to the IBC}
\label{sec:tMAT:IBC}

The MAT scheme (with or without truncation) can always be applied by means of time-sharing among cells, see \mbox{Section \ref{sec:timesharing}}. However, when applied to the IBC with multiple active cells, the interference as of the third phase are no longer totally aligned. To see this, consider as an example a round $r$ of the third phase with $\Ac^{(3,r)} = \{ i,j,k\}$, where $\UE{i}$ and $\UE{j}$ are served by $\BS{1}$, whereas $\UE{k}$ is served by $\BS{2}$. In this case, the property in (\ref{eq:TMAT_RIA_33}) cannot be ensured, since there are multiple transmitters, thus
\begin{IEEEeqnarray}{c}
\RSpan{ 
        \left[ \Hns{i,1}{3,r} \sigmas{3}{3,r} \tng{k,j}{i}  ,  \Hns{i,2}{3,r} \sigmas{3}{3,r} \tng{j,k}{i} \right]} \subseteq \RSpan{ \left[ \tng{k,j}{i} ,  \tng{j,k}{i} \right]},
\end{IEEEeqnarray}
cannot be ensured in general. Conceptually, now the interference is received from two different sources, thus $\tng{k,j}{i}$ and $\tng{j,k}{i}$ travel through different channels, and the interference alignment is broken. Then, this strategy must be updated or extended by taking into account the particular topology of the IBC, as addressed in Section \ref{sec:uMAT}.

\section{Brief Review of the HC Scheme}
\label{sec:HC}

The IBC with a single user per cell ($L=1$) reduces to the $K$-user IC. This section reviews the HC scheme  \cite{Hao_MIMOIC}, which provides the best known DoF inner bound for the MISO IC. Specifically, we review the results for the $K$-user $K \times 1$ IC, i.e. with $K$ antennas at the transmitters and single-antenna receivers.

\subsection{Main Results}

Similarly to the MAT scheme, the achievable DoF are presented in a recursive formulation. Next theorem summarizes the main result:
\\
 
\begin{theorem}[HC scheme, achievable DoF \cite{Hao_MIMOIC}]
For the $K$-user MISO IC with delayed CSIT and $K$ antennas at the transmitters, the following DoF can be achieved:
 \label{th:HC} 
\end{theorem}
\begin{IEEEeqnarray}{c}
 \dHC = \max_{i=1,2} \Bigg( \frac{n_i^2} 
	    { 1 + \frac{n_i (n_i -1)} {\dHC_{2} }} \Bigg).
\label{eq:HC:DoF}
\end{IEEEeqnarray}
with $n_1 = \lfloor 2 \dHC_2 \rfloor$, $n_2 = \lceil 2 \dHC_2 \rceil$, and
\begin{IEEEeqnarray}{c}
 \dHC_2 = \left( 1 - \frac{1}{K-1} \sum_{l=2}^{K-1} \frac{K-l}{l^2 -1}\right)^{-1}.
\label{eq:HC:DoF2}
\end{IEEEeqnarray}

It is worth pointing out that the expression in (\ref{eq:HC:DoF}) does not scale with $K$, thus it collapses to a constant as $ K \rightarrow \infty$.

\subsection{Overview of the Transmission Strategy}
\label{sec:ovHC}

Linear combinations of all the transmitted symbols are sent along $K$ phases. During the first phase, groups of $n$ users are served in orthogonal rounds. This generates order-2 symbols, to be transmitted during phase 2. However, during the second phase, the transmissions to each member of each group (in this case pairs) are orthogonalized in time. The authors call this a 1-Tx/$p$-user scheduling, and again, it generates order-3 symbols.

In contrast to other phases, from phase 3 onwards, each phase $p$ is in turn divided into two sub-phases, namely phases $p$-I and $p$-II. During the first sub-phase, a 1-Tx/$p$-user scheduling is carried out, whereas the second sub-phase follows a similar pattern to the MAT scheme, but with a $p$-Tx/$p$-user scheduling\footnote{Indeed, the MAT scheme employs always a single transmitter, but the number of users whose symbols are transmitted vary depending on the phase number the same way.}.

The main innovation of the HC scheme is the introduction of this first sub-phase, which allows uncoupling the overheard interference at each receiver. As the reader will see later, the MAT extension proposed in this work follows a similar approach.

\subsection{Applicability to the IBC}
\label{sec:HC:IBC}

As for MAT, time-sharing allows reducing the IBC to a $C$-user $L \times 1$ IC. In order to serve all $K=L\cdot C$ users simultaneously, an alternative approach is to consider each transmitter as $L$ different transmitters without antenna cooperation. However, in order to apply the HC scheme for the equivalent $K$-user IC, this would require $L^2$ antennas at each BS. While the MAT and uMAT schemes do not provide additional DoF with more than $L$ antennas at the BSs (see \cite{MAT} and Section \ref{sec:uMAT}), for the sake of comparison we will concede the small privilege of no limitation on the number of antennas at the BSs when simulating the HC scheme. Results in next section show how even with this small advantage, uMAT is superior than HC when applied to the IBC.


\section{Main Results}
\label{sec:MainResults}

The main result of this work is \mbox{Theorem \ref{th:innerbound}}, that provides a DoF inner bound for the IBC with $C$ cells and $L$ users per cell, based on the proposed scheme called \textit{uMAT}. 

\vskip 5mm

\begin{minipage}{\linewidth}
\begin{theorem}[DoF Inner bound]
For the $(L,C)$ MISO IBC with delayed CSIT, and $L$ antennas at the BSs, the following DoF can be achieved:
 \label{th:innerbound} 
\end{theorem}
\def\arraystretch{1.6}
\begin{IEEEeqnarray}{c}
\duMAT = \frac{K}{ \sum\limits_{p=1}^{L} \! \frac{1}{p} + (C-1) 
+ \sum\limits_{p=2}^{L-1} \dfrac{1}{ p \binom{K}{p} } \sum\limits_{r=2}^{\min\left(p, C \right)} 
\binom{C}{r}
\sum\limits_{ \mathbf{g} \in \mathcal{R}_p^r }
\prod\limits_{j =1}^r \binom{L}{g_j} },
\label{eq:uMAT_DoF}
\end{IEEEeqnarray}
where $\mathcal{R}_i^r$ is defined as:
\begin{IEEEeqnarray}{c}
\mathcal{R}_p^r = \Big\{ \left(g_1, \ldots, g_r \right) \,|\, g_1, \ldots, g_r \in \mathbb{Z}^+, \, \sum\limits_{j=1}^r g_j = p \Big\},
\label{eq:Ridef}
\end{IEEEeqnarray}
and only for $i \geq r$.
\end{minipage} 
\vskip 5mm

\begin{IEEEproof}  
See Section \ref{sec:uMAT:Ccells}.
\end{IEEEproof}
\textcolor{white}{res} 

Comparison to previous state-of-the-art (MAT, CSG and HC)\footnote{Our result in \cite{TAV_IBC_ISIT} for the (2, 2) IBC (achieving $\frac{12}{7}$ DoF sum) has been omitted, which would represent just a point in this figure.} and the cooperation upper bound (Coop) is shown in \mbox{Fig. \ref{fig:mainResults}}, where Coop is computed as in (\ref{eq:MATbis}), with $K=L \cdot C$. It is clear that tighter upper bounds are required for this channel.

\begin{figure}
\begin{minipage}{\linewidth}
\begin{minipage}{0.48\linewidth}
\begin{tikzpicture}
\begin{axis}[
  ymin=0,ymax=20,
  xmin=3,xmax=40,
  xmajorgrids,
   ymajorgrids,
   extra y ticks={2.5, 7.5, 12.5},
   grid style={dashed, gray!30},
  ylabel style={at={(0.03,0.5)},rotate=0},
  mark size=1pt,
  width=0.99\linewidth, 
  font=\footnotesize,
  xlabel=Number of users per cell $L$,
  ylabel=DoF sum,
  legend style={at={(axis cs: 3,20)},anchor=north west}],
  ]
  \pgfplotstableread[col sep=comma]{DoFs.csv} \dof
  
      \addplot[plotoptsPropIn5] 
  table[x =L, y =uMAT_C5]   from \dof ; 
  \addlegendentryexpanded{uMAT (C=5)};   
  
  
  \addplot[plotoptsPropIn3] 
  table[x =L, y =uMAT_C3]   from \dof ; 
  \addlegendentryexpanded{uMAT (C=3)};    
   
  \addplot[plotoptsPropIn2] 
  table[x =L, y =uMAT_C2]   from \dof ; 
  \addlegendentryexpanded{uMAT (C=2)};
     
  \addplot[plotoptsCoop3]   
  table[x =L, y =MAT_C3]   from \dof ; 
  \addlegendentryexpanded{Coop (C=3)};
  
  \addplot[plotoptsCoop2]   
  table[x =L, y =MAT_C2]   from \dof ; 
  \addlegendentryexpanded{Coop (C=2)};
  
  \addplot[plotoptsMAT]   
  table[x =L, y =MAT_TDMA]   from \dof ; 
  \addlegendentryexpanded{MAT (C=1)};

  \addplot[plotoptsCastanheiraC5]   
  table[x =L, y =CSG_C5]   from \dof ; 
  \addlegendentryexpanded{CSG (C=5)};
    
  \addplot[plotoptsIC5]   
  table[x =L, y =Hao_C5]   from \dof ; 
  \addlegendentryexpanded{HC (C=5)};
  
  \addplot[plotoptsIC2]   
  table[x =L, y =Hao_C2]   from \dof ; 
  \addlegendentryexpanded{HC (C=2)};

\end{axis}
\end{tikzpicture}
\caption{DoF sum for different number of users per cell and active cells $C$. Proposed scheme uMAT is compared to applying MAT to a single cell (MAT), applying the two best known strategies for the IC (HC and CSG), and the cooperative upper bound (Coop).}
\label{fig:mainResults} 
\end{minipage}
\begin{minipage}{0.04\linewidth}
\hspace{1mm}
\end{minipage}
\begin{minipage}{0.48\linewidth}
\begin{tikzpicture}
\begin{axis}[
  ymin=0,ymax=6,
  xmin=3,xmax=40,
  xmajorgrids,
   ymajorgrids,
      extra y ticks={1, 3, 5},
   grid style={dashed, gray!30},
  ylabel style={at={(0.03,0.5)},rotate=0},
  width=0.99\linewidth, 
    mark size=1pt,
   extra x ticks={},
   extra x tick labels={},
  font=\footnotesize,
  xlabel=Number of users per cell $L$,
  ylabel=$\epsilon$,
  legend style={at={(axis cs: 3,6)},anchor=north west}],
  
  \pgfplotstableread[col sep=comma]{DoFs.csv} \dof
  
  \addplot[plotoptsPropIn5] 
  table[x =L, y expr={\thisrow{uMAT_C5} - \thisrow{MAT_TDMA}}]   from \dof ; 
  \addlegendentryexpanded{$C=5$};

  \addplot[plotoptsPropIn4] 
  table[x =L, y expr={\thisrow{uMAT_C4} - \thisrow{MAT_TDMA}}]   from \dof ; 
  \addlegendentryexpanded{$C=4$};
    
  \addplot[plotoptsPropIn3] 
  table[x =L, y expr={\thisrow{uMAT_C3} - \thisrow{MAT_TDMA}}]   from \dof ; 
  \addlegendentryexpanded{$C=3$};
    
  \addplot[plotoptsPropIn2] 
  table[x =L, y expr={\thisrow{uMAT_C2} - \thisrow{MAT_TDMA}}]   from \dof ; 
  \addlegendentryexpanded{$C=2$};  
  
\end{axis}
\end{tikzpicture}
\caption{Gap (difference between uMAT and MAT) as a function of the number of users $L$ for different number of simultaneously active cells $C$.\textcolor{white}{dasdsadasdsadsaddadasd
dasdsadasdsadsaddadasd
dasdsadasdsadsaddadasd
dasdsadasdsadsaddadasd
dasdsadasdsadsaddadasd
}}
\label{fig:gap} 
\end{minipage}
\end{minipage}
\end{figure}  

As for MAT, the DoF of uMAT scale with $L$, and results in Fig. \ref{fig:mainResults} suggest they scale with $C$ as well. We have not analyzed whether this scaling with $C$ applies indefinitely, but it is an interesting line of future work. Moreover, we observe that the slope of the curve w.r.t. $L$ is larger as $C$ increases. Therefore, to obtain a simple rule of thumb we could linearize these curves and obtain some conclusions. For example, for the MAT scheme we need about 5 more users per cell to obtain an extra DoF, whereas in a scenario with 4 cells uMAT would need approximately 3 more users per cell. Note that in the former case that would add 20 new users to the whole system (5 in each of the 4 cells) compared to 12 new users for the latter case.

While the DoF for uMAT are always greater than those obtained applying the original MAT, the HC scheme outperforms uMAT for $C>2$, $L=3$, even though as commented in Section \ref{sec:HC:IBC}, we are assuming additional antennas for those cases. The CSG scheme never outperforms uMAT for the considered settings. Therefore, as expected, the results in Fig. \ref{fig:mainResults} show that, for most $(L, C)$ settings, having $L$ antennas at each BS cooperating to serve the $L$ users of their cell (uMAT) outperforms having $L^2$ antennas per BS serving in a non-cooperative fashion (HC) the same total number of users. This is an evidence of the power of cooperative transmission over additional non-cooperative antennas.

We also prove the following corollary for the case of $C=2$ cells and conjecture, based on aforementioned results (see \mbox{Fig. \ref{fig:gap}}), its extension to any number of cells $C$:

\vskip 5mm

   \begin{minipage}{\linewidth}
\begin{corollary}[Asymptotic behavior with respect to \cite{MAT} for $C=2$ cells]
For the scenario in Theorem \ref{th:innerbound} with $C=2$ cells:
 \label{coro:gap} 
\end{corollary}
\def\arraystretch{1.6}
\begin{IEEEeqnarray*}{c}
\lim_{L \rightarrow \infty} \epsilon  \rightarrow \infty,
\end{IEEEeqnarray*}
where $\epsilon = \duMAT - \dMAT$, with $\dMAT$ representing the application of MAT with time sharing concepts, i.e. serving only one cell simultaneously.
\end{minipage}
\vskip 5mm

\begin{IEEEproof}  
See Section \ref{sec:uMAT:gap}.
\end{IEEEproof}
\textcolor{white}{res} \\

 \begin{minipage}{\linewidth}
\begin{conjecture}[Asymptotic behavior with respect to \cite{MAT} for $C$ cells]
For the scenario in Theorem \ref{th:innerbound}, and based on our results, we conjecture that the DoF gap as defined in Corollary \ref{coro:gap} holds for any number of cells $C$.
 \label{conjecture} 
\end{conjecture}
\end{minipage}
\vskip 5mm

Finally, the total number of time slots for uMAT are derived. A similar algorithm can be used to obtain the time slots for the HC scheme, and it is omitted to avoid redundancy. The CSG scheme is omitted here as it assumes an infinite number of time slots.

\vskip 5mm

\begin{minipage}{\linewidth}
\begin{corollary}[Total number of time slots]
For the $(L,C)$ MISO IBC, the DoF stated in Theorem \ref{th:innerbound} can be achieved using the uMAT scheme in $\tau$ time slots, see (\ref{eq:tau}), with:
 \label{coro:slots} 
\end{corollary}
\def\arraystretch{1.6}
\begin{subequations} 
\begin{IEEEeqnarray}{c}
R^{\text{uMAT}}_p = \binom{K}{p} + \nu^{\text{uMAT}}_p, \label{eq:RTMAT}\\
S^{\text{uMAT}}_p =  \frac{ \alpha_{L-1}} {\beta_{L-1}} \lambda,
\end{IEEEeqnarray}
\label{eq:tau_uMAT:extradefs}
\end{subequations}
with
\begin{IEEEeqnarray}{c}
\nu^{\text{uMAT}}_p = \left\{ \begin{matrix} 
0 & p=1, L \\
\sum\limits_{r=2}^{\min\left(p, C \right)} \binom{C}{r} \sum\limits_{ \mathbf{g} \in \mathcal{R}_i^r } \prod\limits_{j =1}^r \binom{L}{g_j} & \qquad p=2, ..., L-1
\end{matrix}
\right.,
\label{eq:nu_def}
\end{IEEEeqnarray}
and
\begin{subequations} 
\begin{IEEEeqnarray}{c}
\lambda = \Lcm{ \bbeta_2, \cdots , \bbeta_{L-1} }, \\
\alpha_p = \prod\limits_{j=1}^{p-1} j, \\
 \beta_p = \left\{ \begin{matrix} 
 1 & p=1 \\
 \prod\limits_{j=1}^{p-1} (K - j) & \qquad \qquad p=2, ..., L-1 \\
 \beta_{L-1} & p= L \\
\end{matrix}
\right..
\label{eq:tau_uMAT:beta}
\end{IEEEeqnarray}
\label{eq:tau_uMAT:alpha_beta}
\end{subequations}
\end{minipage} 
\vskip 5mm

\begin{IEEEproof}  
See Section \ref{sec:uMAT:slots}.
\end{IEEEproof}
\textcolor{white}{res} \\

\begin{figure}
\begin{minipage}{\linewidth}
\begin{minipage}{0.48\linewidth}
\begin{tikzpicture}
\begin{semilogxaxis}[
  ymin=1.5, ymax=5,
  xmin=1, xmax=3e17,
  xmajorgrids,
   ymajorgrids,
   grid style={dashed, gray!30},
  ylabel style={at={(0.03,0.5)},rotate=0},
  mark size=1pt,
  width=0.99\linewidth, 
   extra x ticks={},
   extra x tick labels={},
  font=\footnotesize,
  xlabel=Number of time slots $\tau$,
  ylabel=Achievable DoF sum,
  legend style={at={(1,1)},anchor=north east}
  ],
  \pgfplotstableread[col sep=comma]{taus.csv} \dof
  \addplot[plotoptsPropIn2] 
  table[x=tau_uMAT_C2, y =dof_uMAT_C2]   from \dof ; 
  \addlegendentryexpanded{uMAT (C=2)};
%
\addplot[plotoptsMAT]   
  table[x=tau_MATinC2, y =dof_MAT]   from \dof ; 
  \addlegendentryexpanded{MAT (C=1)};
  
  \addplot[plotoptsIC2]   
  table[x=tau_Hao_C2, y =dof_Hao_C2]   from \dof ; 
  \addlegendentryexpanded{HC (C=2)};
\end{semilogxaxis}
\end{tikzpicture}
\end{minipage}
\begin{minipage}{0.04\linewidth}
\hspace{1mm}
\end{minipage}
\begin{minipage}{0.48\linewidth}
\begin{tikzpicture}

\begin{semilogxaxis}[
  ymin=1.5, ymax=5,
  xmin=1, xmax=3e17,
  xmajorgrids,
   ymajorgrids,
   grid style={dashed, gray!30},
  ylabel style={at={(0.03,0.5)},rotate=0},
  mark size=1pt,
  width=0.99\linewidth, 
   extra x ticks={},
   extra x tick labels={},
  font=\footnotesize,
  xlabel=Number of time slots $\tau$,
  ylabel=Achievable DoF sum,
    legend style={at={(1,1)},anchor=north east}
  ],
  \pgfplotstableread[col sep=comma]{taus.csv} \dof
%
    \addplot[plotoptsPropIn4] 
  table[x=tau_uMAT_C4, y =dof_uMAT_C4]   from \dof ; 
  \addlegendentryexpanded{uMAT (C=4)};    
  
  \addplot[plotoptsPropIn3] 
  table[x=tau_uMAT_C3inC4, y =dof_uMAT_C3]   from \dof ; 
  \addlegendentryexpanded{uMAT (C=3)};    
  \addplot[plotoptsPropIn2] 
  table[x=tau_uMAT_C2inC4, y =dof_uMAT_C2]   from \dof ; 
  \addlegendentryexpanded{uMAT (C=2)};
%
  \addplot[plotoptsMAT]   
  table[x=tau_MATinC4, y =dof_MAT]   from \dof ; 
  \addlegendentryexpanded{MAT (C=1)};
  \addplot[plotoptsIC4]   
  table[x=tau_Hao_C4, y =dof_Hao_C4]   from \dof ; 
  \addlegendentryexpanded{HC (C=4)};
  
  \addplot[plotoptsIC3]   
  table[x=tau_Hao_C3inC4, y =dof_Hao_C3]   from \dof ; 
  \addlegendentryexpanded{HC (C=3)};
  
  \addplot[plotoptsIC2]   
  table[x=tau_Hao_C2inC4, y =dof_Hao_C2]   from \dof ; 
  \addlegendentryexpanded{HC (C=2)};
\end{semilogxaxis}
\end{tikzpicture}
\end{minipage}
\end{minipage}
\caption{DoF-delay (Achievable DoF sum vs Number of time slots) for $L=2$ (left) and $L=4$ (right) for uMAT, MAT, and HC, and multiple values of $C$. CSG scheme omitted as it assumes an infinite number of time slots.}
\label{fig:tradeoff}
\end{figure}
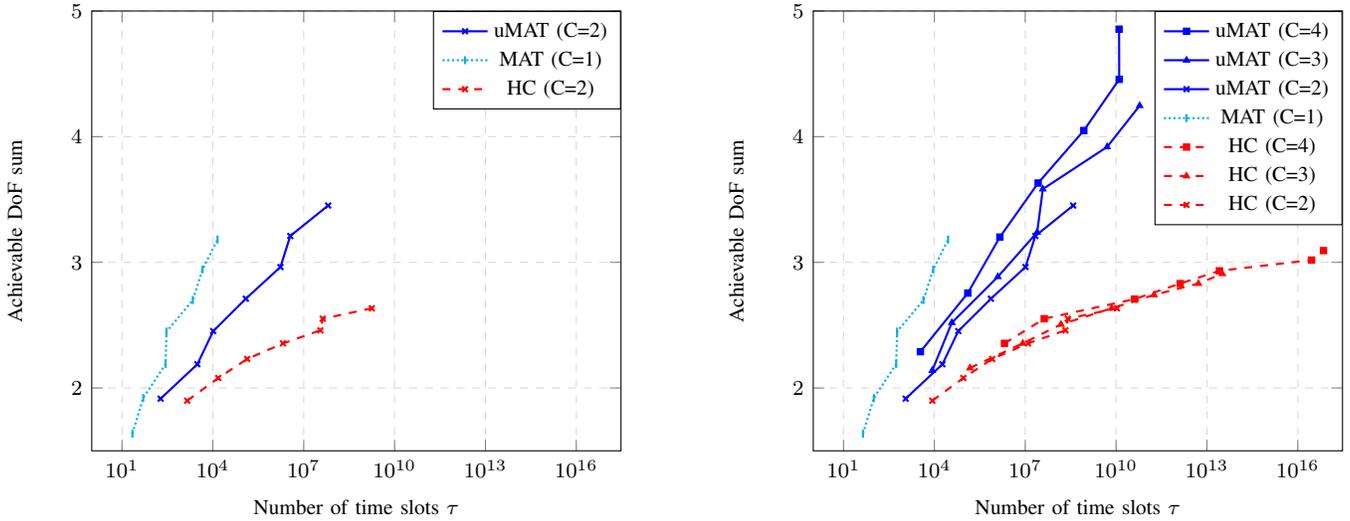

Fig. \ref{fig:tradeoff} shows the comparison to previous state-of-the-art in terms of DoF-delay  (as introduced in \cite{TAV_TIT}) for $L=2$ and $L=4$. As explained before, uMAT provides higher DoF than MAT at the cost of longer transmission delays, whereas compared to HC it is generally superior in both axis.

\section{Uncoupled MAT Scheme}
\label{sec:uMAT}

We propose a low-complexity solution to exploit the MAT scheme philosophy for the IBC. \mbox{Section \ref{sec:uMAT:concept}} introduces the concept powering this extension. Then, the formulation of the DoF achieved using this approach is derived for the case of 2 cells in \mbox{Section \ref{sec:uMAT:2cells}}, with the proof of Corollary \ref{coro:gap} in \mbox{Section \ref{sec:uMAT:gap}}, i.e. how the gap between uMAT and MAT grows indefinitely. Then, the general case for $C$ cells is addressed in \mbox{Section \ref{sec:uMAT:Ccells}}, proving Theorem \ref{th:innerbound}. Finally, Section \ref{sec:uMAT:slots} proves \mbox{Corollary \ref{coro:slots}}, deriving the required time slots to apply this scheme for each setting.

\subsection{Concept}
\label{sec:uMAT:concept}

The problem described in Section \ref{sec:tMAT:IBC} arises from the fact that $\tng{k,j}{i}$ and $\tng{j,k}{i}$ (defined in (\ref{eq:OHI2})) are not known individually, but {\it coupled}. While this is not a problem for the BC, it represents a problem when applying the same transmission scheme to the IBC. In this regard, notice that even in the IBC 
\begin{IEEEeqnarray}{c}
\RSpan{ 
        \left[ \Hns{i,1}{3,r} \sigmas{3}{3,r} \tng{k,j}{i}  ,  \Hns{i,2}{3,r} \sigmas{3}{3,r} \tng{j,k}{i} \right]} \subseteq \RSpan{ \begin{bmatrix} \tng{k,j}{i} & \0 \\ \0 &  \tng{j,k}{i} \end{bmatrix}}
\end{IEEEeqnarray}
can always be ensured, which can be interpreted as knowing the two terms individually, or \textit{uncoupled}. 
While many uncoupling procedures can be elucidated, e.g. the one in \cite{TAV_IBC_ISIT} based on redundancy transmission, this paper proposes the low-complexity solution of \textit{separating in time the transmissions of different transmitters}. Specifically, if one round entails that two or more transmitters are active at the same time, then the round will be repeated, such that the same information is sent by the transmitters, but in different time slots. Specifically, if $p$ BSs are involved in a specific round, that round will have to be repeated $p-1$ times. This very simple solution allows receiving the OHI uncoupled thus satisfying the alignment conditions of the original MAT scheme.
 
In general, the idea is to extend the slots of each round of a phase such that all the OHI is obtained uncoupled. Notice that not all rounds need to be extended, only in case multiple transmitters are active. Since this methodology allows uncoupling the OHI and it is based on the MAT scheme, we label the scheme as  uncoupled MAT or uMAT.

\subsection{Two Active Cells}
\label{sec:uMAT:2cells}

The DoF achieved by the uMAT scheme and presented in \mbox{Theorem \ref{th:innerbound}} are first derived for the $(L,2)$ IBC. As a building block for the proof of the theorem, consider the following lemma:

\vskip 5mm

\begin{minipage}{\linewidth}
\begin{lemma}[Order-$p$ DoF inner bound for $C=2$ cells]
For the $(L, 2)$ MISO IBC with delayed CSIT, and $L$ antennas at the two BSs, the following order-$p$ DoF, $p= 2,\dots, L-1$ , can be achieved:
 \label{lem:2C} 
\end{lemma}
\def\arraystretch{1.6}
\begin{subequations} 
\begin{IEEEeqnarray}{rl}
 \duMAT_p & = \frac{(2L-p+1) \binom{2L}{p}} 
	    { 2\left( \binom{2L}{p} - \binom{L}{p} \right) + \frac{p \binom{2L}{p+1}} {\duMAT_{m+1} }}   \label{eq:lem2:easy} \\
	     & = \frac{(2L-p+1) \binom{2L}{p}} 
	    { \binom{2L}{p} + \!\! \sum\limits_{ \mathbf{g} \in \mathcal{R}_p^2 } \!\!
	     \binom{L}{g_1}\binom{L}{g_2} +
	    \frac{p \binom{2L}{p+1}} {\duMAT_{p+1} }},
\label{eq:lem2:general}
\end{IEEEeqnarray}
\label{eq:lem2:group}
\end{subequations}
with $\mathcal{R}_p^2$ as defined in (\ref{eq:Ridef}).
\end{minipage} 
\vskip 5mm

\begin{IEEEproof}  
This DoF expression is very similar to the one for the MAT scheme, see (\ref{eq:MAT:DoFp}). However, among the $\binom{K}{p} = \binom{2L}{p}$ total number of groups of $p$ users, the transmission should be repeated in case the two transmitters are simultaneously active. Since in each cell there are $\binom{L}{p}$ groups of $p$ users, the number of repetitions is equal to 
\begin{IEEEeqnarray}{c}
\nu^{\text{uMAT}}_p = \binom{2L}{p} - 2\binom{L}{p}. 
\label{nuC2}
\end{IEEEeqnarray}
Adding this extra number of rounds into the denominator of (\ref{eq:MAT:DoFp}), one ends up with (\ref{eq:lem2:easy}).

With a view towards generalization to more cells, we will show that $\duMAT$ can also be written as (\ref{eq:lem2:general}). To this end, recall on the set $\mathcal{R}_p^r$, defined in (\ref{eq:Ridef}) and rewritten here for reader's convenience for $r=2$:
\begin{IEEEeqnarray*}{c}
\mathcal{R}_p^2 = \left\{ \left(g_1, g_2 \right) \,|\, g_1, g_2 \in \mathbb{Z}^+, \, g_1 + g_2 = p \right\}. 
\end{IEEEeqnarray*}
The elements of this set can be interpreted as the number of users selected from each cell, with at least one per cell. In other words, $g_1$ and $g_2$ are the number of users selected from each of the two cells, respectively. Therefore, (\ref{nuC2}) can be rewritten as
\begin{IEEEeqnarray*}{c}
\nu^{\text{uMAT}}_p = \sum\limits_{ \mathbf{g} \in \mathcal{R}_p^2 } \!\!\!\binom{L}{g_1} \binom{L}{g_2}.
\end{IEEEeqnarray*}

Now, notice that during the first phase of the MAT scheme only one transmitter is simultaneously active. Hence, the first phase is unaltered. Then, considering a transmission truncated to $L$ phases\footnote{It has been observed by simulation that more than $L$ phases make decrease the achieved DoF.} means that we force
\begin{IEEEeqnarray}{c}
\duMAT_L = 1.
\label{eq:2cells_f1}
\end{IEEEeqnarray}
Using $\duMAT_1 = \dMAT_1$ as defined in (\ref{eq:MAT:DoFp}), and solving the recursive equation, one ends up with
\begin{IEEEeqnarray}{c}
\duMAT = \frac{L}{\sum\limits_{p=1}^{L} \! \frac{1}{p} - \frac{1}{2L} - \sum\limits_{p=2}^{L-1} \! \Big( \frac{1}{p} \prod\limits_{j=0}^{p-1} \frac{L-j}{2L-j} \Big)}.
\label{eq:uMAT:C2}
\end{IEEEeqnarray}
equal to (\ref{eq:uMAT_DoF}) for $C=2$, thus the proof is complete.\\
\end{IEEEproof}

\subsection{DoF gap for Two Active Cells (Proof of Corollary \ref{coro:gap})}
\label{sec:uMAT:gap}

Based on (\ref{eq:MAT}) and (\ref{eq:uMAT:C2}), the DoF gap is defined as
\begin{IEEEeqnarray}{c}
\epsilon = \frac{L}{\sum\limits_{p=1}^{L} \! \frac{1}{p} - \frac{1}{2L} - \kappa} - \frac{L}{\sum\limits_{i=1}^{L} \! \frac{1}{p}} = L \frac{\kappa + \frac{1}{2L}}{\left(\sum\limits_{p=1}^{L} \! \frac{1}{p} + \frac{1}{2L} - \kappa \right) \left( \sum\limits_{p=1}^{L} \! \frac{1}{p} \right)}.
\label{gap}
\end{IEEEeqnarray}
where $\kappa = \sum\limits_{p=2}^{L-1} \! \Big( \frac{1}{p} \prod\limits_{j=0}^{p-1} \frac{L-j}{2L-j} \Big)$, with the following asymptotic behavior:
\begin{IEEEeqnarray*}{c}
\kappa = \sum\limits_{p=2}^{L-1} \! \Big( \frac{1}{p} \prod\limits_{j=0}^{p-1} \frac{L-j}{2L-j} \Big) \leq \sum\limits_{p=2}^{L-1} \! \Big( \frac{1}{p} \prod\limits_{j=0}^{p-1} \frac{1}{2} \Big) =
\sum\limits_{p=1}^{L-1} \! \frac{1}{p} \left( \frac{1}{2} \right)^p - \frac{1}{2} \overset{L \rightarrow \infty}{=} \log2 - \frac{1}{2}.
\end{IEEEeqnarray*}
According to this, the proof follows:
\begin{IEEEeqnarray*}{c}
\lim_{L \rightarrow \infty} \epsilon  =
\lim_{L \rightarrow \infty} L \frac{\kappa + \frac{1}{2L}}{\left(\sum\limits_{p=1}^{L} \! \frac{1}{p} + \frac{1}{2L} - \kappa \right) \left( \sum\limits_{p=1}^{L} \! \frac{1}{p} \right)} = 
\lim_{L \rightarrow \infty} \frac{L\kappa}{\left( \log L \right)^2},
\end{IEEEeqnarray*}
thus the gap increases indefinitely as $L$ grows, and the proof is complete.

\subsection{General case: $C$ Cells (Proof of Theorem \ref{th:innerbound})}
\label{sec:uMAT:Ccells}

The proof is analogous to the case of 2 cells. Consider the following lemma:

\vskip 5mm

\begin{minipage}{\linewidth}
\begin{lemma}[Order-$p$ DoF inner bound for $C$ cells]
For the $(L, C)$ MISO IBC with delayed CSIT, and $L$ antennas at the BSs, the following order-$p$ DoF can be achieved:
 \label{lem:C} 
\end{lemma}
\def\arraystretch{1.6}
\begin{IEEEeqnarray}{c}
 \duMAT_p = \frac{(K-p+1) \binom{K}{p}} 
	    { \binom{K}{p} + \!\!\! \sum\limits_{r=2}^{\min\left(p, C \right)}  \!\!\!
	      \binom{C}{r} \sum\limits_{ \mathbf{g} \in \mathcal{R}_p^r }
				  \prod\limits_{j =1}^r \binom{L}{g_j} +
	    \frac{p \binom{K}{p+1}} {\duMAT_{p+1} }}, 
	    \quad p= 2,\dots, L-1.
\label{lem:C}
\end{IEEEeqnarray}
\end{minipage} 
\vskip 5mm

\begin{IEEEproof}  
This DoF expression is analogous to (\ref{eq:lem2:general}). Similarly, the second term in the denominator denotes the repetitions for phase $p$, and follows the definition of $\nu^{(\text{uMAT})}_p$ in (\ref{eq:nu_def}), here repeated for reader's convenience:
\begin{IEEEeqnarray*}{c}
\nu^{\text{uMAT}}_p = \!\!\! \sum\limits_{r=2}^{\min\left(p, C \right)} \!\! \binom{C}{r} \sum\limits_{ \mathbf{g} \in \mathcal{R}_p^r } \prod\limits_{j =1}^r \binom{L}{g_j}, \quad p=2, \dots , L-1
\end{IEEEeqnarray*}
In contrast to the $C=2$ case, in the general case the BSs need to repeat a round up to $C-1$ times (the worst case is when the group involves users from all the possible $C$ cells). Hence, we have a summation from the second transmission (first repetition) to $p$, unless the number of cells $C$ is lower than $p$, thus the $\min(p, C)$. That is because the number of cells denote the maximum number of transmitters.

Another difference w.r.t. (\ref{eq:lem2:general}) is that the groups with multiple transmitters need to be accounted for all possible combinations of $r$ cells, hence the term $\binom{C}{r}$. Finally, notice that the previous product $\binom{L}{g_1}\binom{L}{g_2}$ is now generalized to a product of $r$ terms, where $g_j$ denotes the $j$th element of $\mathbf{g}$.
\end{IEEEproof}

\vskip 5mm

It is worth pointing out that $\nu^{(\text{uMAT})}_p=0$ for $p=1,L$, as no repetitions are required for both the first and the last phase. On the one hand, during the first phase rounds there is only one transmitter active at a time. Then, there is no need to repeat any round as interference is already uncoupled. On the other hand, for phase $L$ some of the interference is certainly received coupled, but there are no phases that would benefit from uncoupling it, thus no rounds are repeated.

Solving the recursive equation in (\ref{lem:C}) and using same expression for $\duMAT_1$ as for $\dMAT_1$ in (\ref{eq:MAT:DoFp}), one ends up with (\ref{eq:uMAT_DoF}), and the proof of Theorem \ref{th:innerbound} is complete.

\subsection{Number of time slots (Proof of Corollary \ref{coro:slots})}
\label{sec:uMAT:slots}

The recursive equation in (\ref{lem:C}) can be written by using the definitions in (\ref{eq:tau_uMAT:extradefs})-(\ref{eq:nu_def}) as
\begin{IEEEeqnarray}{c}
 \duMAT_p = \frac{(K-p+1) \binom{K}{p}} 
	    { R^{(\text{uMAT})}_p +
	    \frac{p \binom{K}{p+1}} {\duMAT_{p+1} }}, 
	    \, p= 2,\dots, L-1,
\label{lem:C:nu}
\end{IEEEeqnarray}
where $R^{(\text{uMAT})}_p$ includes the repetitions for phase $p$, see (\ref{eq:RTMAT}). Consider the unfolding of the recursive equation in (\ref{lem:C:nu}), but keeping the original numerator, and without simplifying non-equal terms\footnote{If non-equal terms were simplified, some of the required rounds would be simplified, and this method would not be valid to derive the number of times slots.}. One ends up with the following expression:
\begin{IEEEeqnarray}{c}
 \duMAT_1 = \frac{K^2} 
	    { R^{(\text{uMAT})}_1 + \sum\limits_{p=2}^{L-1} 
	    R^{\text{(uMAT)}}_p \prod\limits_{j=1}^{p-1} \frac{j}{K-j}
	      + 
	    R^{\text{(uMAT)}}_L \left( L-1 \right) \prod\limits_{j=1}^{L-2} \frac{j}{K-j}  	
	    	 },
\label{lem:C:before_lambda}
\end{IEEEeqnarray}
where we can identify the definitions of $\alpha_i, \beta_i$, see (\ref{eq:tau_uMAT:alpha_beta}). Now, if we use the notation $\bar\alpha_{p}, \bar\beta_{p}$ as defined in (\ref{eq:gcd}), (\ref{lem:C:before_lambda}) is simplified to
\begin{IEEEeqnarray}{c}
 \duMAT_1 = \frac{K^2} 
	    { R^{(\text{uMAT})}_1 + \sum\limits_{p=2}^{L-1} \frac{ \bar\alpha_{p}} {\bar\beta_{p}} R^{\text{(uMAT)}}_p  + \frac{\bar\alpha_{L}} {\bar\beta_{L}} R^{\text{(uMAT)}}_L }.
\label{lem:C:with_alpha_beta}
\end{IEEEeqnarray}
As the DoF can be seen as the number of symbols transmitted divided by the number of times slots (see (\ref{eq:achDoF})), our objective now is to see what is the minimum factor, say $\lambda$, to make the denominator of (\ref{lem:C:with_alpha_beta}) equal to an integer number. Taking into account that $R^{\text{(uMAT)}}_p$ is an integer value for all $p$, and $\beta_{L} = \beta_{L-1}$, see (\ref{eq:tau_uMAT:beta}), then it reduces to
\begin{IEEEeqnarray}{c}
\lambda = \Lcm{ \bbeta_2, \cdots , \bbeta_{L-1} },
\label{lem:C:lambda_again}
\end{IEEEeqnarray}
and this completes the proof.

\section{Conclusion}
\label{sec:conclusion}
 
A low complexity adaptation of the MAT scheme for the BC has been proposed for the IBC with $L$ users per cell and $C$ cells, denoted as uncoupled MAT or uMAT scheme. Compared to the straightforward application of MAT to the IBC via time sharing, it achieves higher DoF at the cost of higher transmission delays. The DoF gain is, however, notable, and it will depend on the affordable system complexity. 

The key concept behind uMAT is the \textit{interference uncoupling}. In this work, the simplest approach has been exploited, i.e. orthogonalization in time for every transmission of symbols involving multiple transmitters. However, as already proposed for the IC, e.g. HC scheme in \cite{Hao_MIMOIC}, or in \cite{TAV_IBC_ISIT} for the IBC, there are probably better ways to deal with this problem for particular settings. The optimization of such uncoupling mechanism remains as future work.

Additionally, there are a number of topics that remain unsolved after this work. First, if our focus is in optimizing the DoF at any complexity cost, a promising research direction would be using a scheme of $K= L \cdot C$ phases. Second, it would be useful to derive upper bounds tighter than those based on cooperation so as we can conclude to what extent the performance obtained by uMAT is close to the optimal. And finally, it could be interesting to study other antenna settings for this scenario beyond the MISO case. In this regard, the concept of \textit{IA at the receiver side} introduced in \cite{CastanheiraIC} for the IC (CSG scheme), might be a promising starting point.




\ifCLASSOPTIONcaptionsoff
  \newpage
\fi

\bibliographystyle{IEEEbib}
\bibliography{../../../papers/_referenciesMarc}

\end{document}